\documentclass[11pt,twoside,a4paper,cmspaper,final,collab]{cms-tdr}
\def\svnVersion{31c10b8-D}\def\svnDate{2026/08/10}\def\cmsCernNoTag{CERN-EP-2026-168}\def\cmsCernDate{\today}\def\cmsMessage{Submitted to the European Physical Journal C}
\begin{document}\cmsNoteHeader{B2G-25-005}

\cmsNoteHeader{B2G-25-005}

\newlength\cmsFigWidth
\ifthenelse{\boolean{cms@external}}{\setlength\cmsFigWidth{0.85\columnwidth}}{\setlength\cmsFigWidth{0.4\textwidth}}

\ifthenelse{\boolean{cms@external}}{\providecommand{\cmsTable}[1]{\resizebox{\linewidth}{!}{#1}}}{\providecommand{\cmsTable}[1]{#1}}
\ifthenelse{\boolean{cms@external}}{\providecommand{\onecolnewpage}{}}{\providecommand{\onecolnewpage}{\newpage}}

\newlength\cmsTabSkip\setlength{\cmsTabSkip}{1ex}

\newcommand{\pp}{\ensuremath{\Pp\Pp}\xspace}
\providecommand{\PAQqpr}{\HepParticle{\PAQq}{}{\prime}\xspace}
\newcommand{\ct}{\ensuremath{g_{\PQt\PAQt\PZpr}}\xspace}
\newcommand{\partV}{\PZpr}
\newcommand{\partS}{{\HepParticle{\upphi}{}{}}\xspace}
\newcommand{\partP}{{\HepParticle{a}{}{}}\xspace}
\newcommand{\ttV}{\ensuremath{\ttbar\partV}\xspace}
\newcommand{\ttS}{\ensuremath{\ttbar\partS}\xspace}
\newcommand{\ttP}{\ensuremath{\ttbar\partP}\xspace}

\newcommand{\jetAK}{\ensuremath{\mathrm{AK4}}\xspace}
\newcommand{\jetJ}{\ensuremath{\mathrm{VR}}\xspace}
\newcommand{\tlept}{\ensuremath{\PQt_{\text{lept}}}\xspace}
\newcommand{\thad}{\ensuremath{\PQt_{\text{had}}}\xspace}

\newcommand{\ttZ}{\ensuremath{\PQt\PAQt\PZ}\xspace}
\newcommand{\ttW}{\ensuremath{\PQt\PAQt\PW}\xspace}
\newcommand{\ttH}{\ensuremath{\PQt\PAQt\PH}\xspace}
\newcommand{\VV}{\ensuremath{\PV\PV}\xspace}
\newcommand{\zjets}{\ensuremath{\PZ/\PGgst\text{+jets}}\xspace}
\newcommand{\fourt}{\ensuremath{4\PQt}\xspace}
\newcommand{\threet}{\ensuremath{\PQt\PQt\PQt\PX}\xspace}

\newcommand{\etat}{{\HepParticle{\PGh}{\PQt}{}}\xspace}

\newcommand{\mV}{\ensuremath{m_{\partV}}\xspace}
\newcommand{\mS}{\ensuremath{m_{\partS}}\xspace}
\newcommand{\mP}{\ensuremath{m_{\partP}}\xspace}

\newcommand{\GammaV}{\ensuremath{\Gamma_{\partV}}\xspace}
\newcommand{\GammaS}{\ensuremath{\Gamma_{\partS}}\xspace}
\newcommand{\GammaP}{\ensuremath{\Gamma_{\partP}}\xspace}

\newcommand{\mJJ}{\ensuremath{m_{\mathrm{JJ}}}\xspace}
\newcommand{\mll}{\ensuremath{m_{\Pell\Pell}}\xspace}

\newcommand{\deepjet}{\textsc{DeepJet}\xspace}
\newcommand{\hotvr}{\textsc{HOTVR}\xspace}

\title{Search for a resonance in events with four top quarks decaying into two leptons and jets in proton-proton collisions}

\date{\today}

\abstract{
A search for a heavy resonance in events containing four top quarks is presented and interpreted in models featuring a new vector, scalar, or pseudoscalar boson that interacts exclusively with top quarks. Proton-proton collision data collected by the CMS experiment at the CERN LHC at centre-of-mass energies of 13 and 13.6\TeV are analysed, corresponding to integrated luminosities of 138 and 35\fbinv, respectively. Events with two leptons (electrons or muons) and jets are selected. The mass of the heavy resonance is reconstructed from a pair of jets that are formed using a variable-radius jet clustering algorithm. A machine-learning approach is used to identify whether these jets originate from top quarks that decay fully hadronically. The dominant background from misidentified jets is estimated from data as a function of the dijet invariant mass. No significant excess is observed in data. Upper limits at 95\% confidence level are set on the resonance production cross section for resonance masses between 500\GeV and 4\TeV, and relative widths ranging from 4 to 50\%. In the vector-mediator benchmark, resonances with 50\% relative width are excluded up to masses of 850\GeV, with an expected exclusion limit of 1000\GeV. This is the first dedicated search for these models in the two lepton channel.
}

\hypersetup{%
pdfauthor={CMS Collaboration},%
pdftitle={Search for a resonance in events with four top quarks decaying into two leptons and jets in proton-proton collisions},%
pdfsubject={CMS},%
pdfkeywords={CMS, BSM particle, top quarks, machine learning}}

\maketitle

\section{Introduction}

The observation of the production of four top quarks (\PQt) in proton-proton (\pp) collisions by the ATLAS and CMS Collaborations~\cite{ATLAS:2023ajo,CMS:2023ftu} marks a significant milestone in testing the standard model (SM) of particle physics.
The SM has proved a very successful theory but has known shortcomings, including the lack of an explanation of the nature of dark matter~\cite{Cirelli:2024ssz,Bertone:2004pz,Clowe:2006eq}.
Many theories beyond the SM (BSM) propose new interactions and particles to address its shortcomings, but the ongoing absence of clear deviations from SM predictions places tight constraints on any BSM effects at LHC energy scales. 
These constraints are generally weaker for BSM models which feature new particles that interact dominantly or exclusively with top quarks;
the large mass of the top quark, which results in a Yukawa coupling to the Higgs boson (\PH) close to unity, also hints that it could play a central role in BSM physics~\cite{PhysRevD.20.2619}.
Rare four-top quark events are therefore an interesting probe for such scenarios, especially as measured cross sections are slightly higher than the SM prediction, consistently across all measured final states~\cite{CMS:2023zdh,ATLAS:2021kqb}.

{\tolerance=500
This paper focuses on simplified BSM scenarios that introduce a new vector mediator particle (\partV) that couples exclusively to top quarks at tree level -- so-called top-philic models~\cite{Greiner:2014qna, Kim:2016plm}. These scenarios are chosen as a benchmark to guide the optimisation of the analysis strategy because of their representative phenomenology. For comparison, alternative interpretations involving a scalar (\partS) or pseudoscalar (\partP) mediator---motivated by two-Higgs-doublet models~\cite{Branco:2011iw,PhysRevD.8.1226} or models with an axion-like particle (ALP)~\cite{PhysRevLett.38.1440,PhysRevD.16.1791,PhysRevLett.40.279,Blasi:2023hvb}---are also considered. All of these mediators can also appear as mediators for the production of dark matter~\cite{Cox:2015afa,LHCDMWG:2018ufk,Bharucha:2022lty}.
\par}

The interaction of the \partV boson with the top quark is described by the Lagrangian~\cite{Greiner:2014qna}
\begin{equation}
	\mathcal{L} = \ct\PAQt\gamma_{\mu}\left(\cos\theta P_{\mathrm{L}}+\sin\theta P_{\mathrm{R}}\right)\PQt\partV^{\mu}+\text{h.c.},
	\label{eq:intro-zprime}
\end{equation}
{\tolerance=500
where \ct denotes the net interaction strength and $\theta$ controls the relative interaction strength to left- and right-handed top quarks through the application of the projection operators $P_{\mathrm{L}}$ and $P_{\mathrm{R}}$, respectively. This model also predicts signatures with a single top quark-antiquark pair (\ttbar) through loop-induced \partV production via virtual top quarks, $\pp\to\partV\to\ttbar$. Dedicated searches focusing on \partV boson production in \ttbar events have so far not revealed any evidence for scenarios where the \partV boson mass (\mV) is significantly above the \ttbar production threshold~\cite{ATLAS:2020lks,ATLAS:2024vxm,CMS:2025dzq,CMS:2026lsc}. Therefore, in this search we focus on the case of an equal coupling strength for left- and right-handed top quarks, \ie $\theta=\pi/4$, which maximally suppresses this loop-induced production.
\par}

For the scalar or pseudoscalar mediators, the coupling to top quarks is described as
\begin{equation}
    \mathcal{L}_{\partS} = -g_{\ttS}\frac{m_{\PQt}}{v}\ttS,
    \qquad
    \mathcal{L}_{\partP} = ig_{\ttP}\frac{m_{\PQt}}{v}\PAQt\gamma_{5}\PQt\partP,
\end{equation}
where $v=246\GeV$ is the vacuum expectation value of the SM Higgs field and $g_{\ttS/\ttP}$ is the coupling constant for the scalar or pseudoscalar mediator, respectively. The pseudoscalar mediator could also be an ALP, \PSA, in which case the coupling to top quarks arises via a dimension-5 operator which can be re-expressed in terms of the ALP decay constant $f_{\PSA}$, and a dimension-4 operator~\cite{Blasi:2023hvb},
\begin{equation}
    \mathcal{L}_{\PSA} = ic_{\PQt}\frac{m_{\PQt}}{f_{\PSA}}\PAQt\gamma_{5}\PQt\PSA,
\end{equation}
where $c_{\PQt}$ is the coupling strength to top quarks. This interpretation is of interest as different measurements can exclude different parts of the ALP parameter space, and the result of the analysis presented here is compared with limits from a \ttbar resonance search~\cite{CMS:2025dzq} and a search for a dark matter candidate in association with top quarks~\cite{CMS:2025ncs}.

Dedicated searches for top-philic heavy resonances in events with four top quarks have been conducted by the ATLAS and CMS Collaborations in final states where one top quark decays leptonically, $\PQt \to \PQb\Pell\PAGn$~\cite{ATLAS:2023taw,CMS:B2G-24-009}.
These searches focus on the case of $\mV>1\TeV$, where the top quarks produced in the decay of the \partV boson are significantly Lorentz boosted. 
This results in collimated decay products that can be reconstructed in a single jet when the top quark decays fully hadronically, \ie through the decay chain $\PQt\to\PQb\PW\to\PQb\PQq\PAQqpr$, where \PQq and $\PAQqpr$ denote \PQu, \PQd, \PQc, and \PQs quarks and antiquarks, respectively.

\begin{figure}[htb]
\centering
\includegraphics[width=0.42\textwidth]{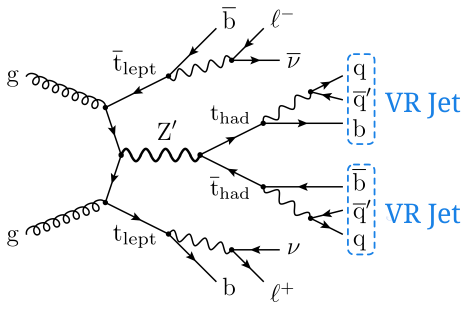}
\caption{\label{fig:selection-signature} The targeted signal signature of \partV boson production. It comprises two leptonically decaying top quarks (\tlept) and two hadronically decaying top quarks (\thad), where the latter two are reconstructed from jets using a variable-radius clustering algorithm.}
\end{figure}

The search presented in this article focuses on a new final state compared to previous searches, requiring two leptons (electrons or muons) with opposite electric charges. Specifically, it targets the topology where the two top quarks produced in association with the \partV boson decay leptonically, and the two top quarks from the mediator decay fully hadronically, as shown in Fig.~\ref{fig:selection-signature}. The \ttS and \ttP processes yield the same topology, with the \partV boson replaced with a mediator \partS or \partP, respectively. We analyse \pp collision data at $\sqrt{s}=13\TeV$ (2016--2018) and $13.6\TeV$ (2022), corresponding to an integrated luminosity of 138\fbinv and 35\fbinv, respectively.
A novelty is the reconstruction of \mV from a pair of jets formed using a variable-radius jet clustering algorithm (\jetJ jets)~\cite{Lapsien:2016zor,CMS:2020poo}, which extends the sensitivity to mediators with masses less than 1\TeV. Additionally, a machine-learning classifier is employed to identify whether the resulting jets originate from genuine top quark decays in the all-hadronic channel. The rate of background processes is estimated from data using a sideband region. Limits on the \ttV, \ttS, and \ttP production cross sections are obtained for mediator widths of 4, 10, 20, and 50\% relative to its mass, in the mass range 500\GeV to 4\TeV. Tabulated results are provided in the HEPData record for this analysis~\cite{hepdata}.

This paper is organised as follows: A brief description of the CMS detector and the reconstruction of relevant physics objects is given in Section~\ref{sec:cms}. Details on the data sets and samples of simulated events are given in Section~\ref{sec:samples}. Section~\ref{sec:selection} describes the selection and categorisation of events for analysis. The determination of the background from data is given in Section~\ref{sec:background}, which is followed by details on the considered sources of systematic uncertainties in Section~\ref{sec:sys}. The results are presented in Section~\ref{sec:results}, and a summary can be found in Section~\ref{sec:summary}. 

\section{The CMS detector and event reconstruction}
\label{sec:cms}

The CMS apparatus~\cite{CMS:2008xjf,CMS:2023gfb} is a multipurpose, nearly hermetic detector, designed to trigger on~\cite{CMS:2020cmk,CMS:2016ngn,CMS:2024aqx} and identify electrons, muons, photons, and (charged and neutral) hadrons~\cite{CMS:2020uim,CMS:2018rym,CMS:2014pgm}. Its central feature is a superconducting solenoid of 6\unit{m} internal diameter, providing a magnetic field of 3.8\unit{T}. Within the solenoid volume are a silicon pixel and strip tracker, a lead tungstate crystal electromagnetic calorimeter (ECAL), and a brass and scintillator hadron calorimeter (HCAL), each composed of a barrel and two endcap sections. Forward calorimeters extend the pseudorapidity coverage provided by the barrel and endcap detectors. Muons are reconstructed using gas-ionisation detectors interleaved with the layers of the steel flux-return yoke outside the solenoid. More detailed descriptions of the CMS detector, together with a definition of the coordinate system used and the relevant kinematic variables, can be found in Refs.~\cite{CMS:2008xjf,CMS:2023gfb}.

Events of interest are selected using a two-tiered trigger system. The first level (L1), composed of custom hardware processors, uses information from the calorimeters and muon detectors to select events at a rate of around 100\unit{kHz} within a fixed latency of 4\mus~\cite{CMS:2020cmk}. The second level, known as the high-level trigger, consists of a farm of processors running a version of the full event reconstruction software optimised for fast processing, and reduces the event rate to a few kHz before data storage~\cite{CMS:2016ngn,CMS:2024aqx}. 

A particle-flow algorithm~\cite{CMS:2017yfk} aims to reconstruct and identify each individual particle in an event, with an optimised combination of information from the various elements of the CMS detector. The energy of photons is obtained from the ECAL measurement. The energy of an electron is determined from a combination of the electron momentum at the primary interaction vertex as determined by the tracker, the energy of the corresponding ECAL cluster, and the energy sum of all bremsstrahlung photons spatially compatible with originating from the electron track. The energy of a muon is obtained from the curvature of the corresponding track. The energy of a charged hadron is determined from a combination of the momentum measured in the tracker and the matching ECAL and HCAL energy deposits, corrected for the response function of the calorimeters to hadronic showers. Finally, the energy of neutral hadrons is obtained from the corresponding corrected ECAL and HCAL energies.

{\tolerance=1000
Standard fixed-radius jets (\jetAK) are reconstructed by clustering PF candidates using the anti-\kt algorithm~\cite{Cacciari:2008gp}, with a distance parameter of $\Delta R = \sqrt{\smash[b]{(\Delta\eta)^2+(\Delta\phi)^2}}=0.4$ (where $\eta$ is the pseudorapidity and $\phi$ the azimuthal angle in radians), as implemented in the \FASTJET package~\cite{Cacciari:2011ma}. In the 2016--2018 data-taking periods, the impact from additional \pp interactions within the same or nearby bunch crossings (pileup) is mitigated using the ``charged-hadron subtraction'' algorithm~\cite{CMS:2016lmd}. The momentum of each jet is defined as the vectorial momentum sum over all its constituent PF candidates and an offset correction is applied to account for remaining pileup contributions. The ``pileup-per-particle identification'' algorithm~\cite{Sirunyan:2020foa,Bertolini:2014bba} is used in the 2022 data-taking period to mitigate the effect of pileup at the reconstructed-particle level, making use of local shape information, event pileup properties, and tracking information.
\par}

Jet energy corrections are derived from simulation so that their average measured energy becomes identical to that of particle-level jets. In situ measurements of the momentum balance in dijet, photon+jet, \PZ+jet, and multijet events are used to determine any residual differences between the jet energy scale in data and in simulation, and appropriate corrections are made~\cite{CMS:2016lmd}. Additional selection criteria are applied to each jet to remove jets potentially dominated by anomalous contributions from various subdetector components or reconstruction failures~\cite{CMS:2017wyc}.

The missing transverse momentum vector \ptvecmiss is computed as the negative vector sum of the transverse momenta (\pt) of all the PF candidates in an event, and its magnitude is denoted as \ptmiss~\cite{CMS:2019ctu}. The \ptvecmiss is modified to account for corrections to the energy scale of the reconstructed jets in the event. Anomalous high-\ptmiss events can arise from a variety of reconstruction failures, detector malfunctions, or non-collision backgrounds. Such events are rejected by event filters that are designed to identify more than 85--90\% of the spurious high-\ptmiss events with a mistagging rate less than 0.1\%~\cite{CMS:2019ctu}.

The primary vertex is taken to be the vertex corresponding to the hardest scattering in the event, evaluated using tracking information alone, as described in Section 9.4.1 of Ref.~\cite{CMS-TDR-15-02}.

\section{Data set and simulated samples}
\label{sec:samples}

The analysed samples of \pp collision data at $\sqrt{s}=13$ and 13.6\TeV were recorded with the CMS detector during 2016--2018 and 2022, and correspond to integrated luminosities of 138 and 35\fbinv, respectively. 
The data sets are analysed independently for each data-taking period and only combined for the final result to account for differences related to the data-taking conditions, and to apply proper corrections and calibrations. These data-taking periods represent periods of similar detector conditions and mainly correspond to calendar years. However, the 2016 data are split into periods before and after the tracker readout chip bias voltage was increased to reduce dead time caused by the higher luminosity (denoted pre- and post-VFP), and the 2022 data are split into periods before and after an intervention to repair a water leak in the ECAL endcap detector (pre- and post-EE).

Events are recorded using a combination of dilepton and single-lepton triggers, where the latter improve the overall trigger efficiency by selecting most events where one lepton fails the trigger-level selections. This combination of triggers is found to be over 98\% efficient for \ttbar events passing the two lepton selection described in Section~\ref{sec:selection}.

Samples of simulated signal events are used to evaluate the detector resolution, efficiency, and acceptance for the signal processes. 
Additionally, samples of simulated background events are used for understanding the composition of the background, deriving the transfer factors used in the estimation of the dominant background from data in sidebands, and predicting minor backgrounds. 
Samples of simulated events are also used to train and characterise machine-learning algorithms.

{\tolerance=600
Samples of \partV boson production in association with two top quarks, $\pp\to\ttV$, with various values of \mV between 500\GeV and 4\TeV, and with relative decay widths $\GammaV/\mV$ of 4, 10, 20, and 50\%, are generated in the five-flavour scheme at leading order (LO) precision using the \MGvATNLO{} v2.9.18 event generator~\cite{Alwall:2014hca}, with the mediator then decaying as $\partV\to\ttbar$. This procedure retains only diagrams in which the \partV mediator is produced in the $s$-channel, and neglects interference with SM four top quark production. The omitted contributions are model dependent and can be substantial, particularly for large mediator masses and widths. However, they are expected to contribute predominantly close to the four top quark production threshold, where the parton luminosities are largest, rather than enhancing the resonant contribution targeted in this search.
Subsequent decays of the top quarks are simulated using \PYTHIA{} v8.240 (2016--2018) or v8.306 (2022). 
Cross sections are computed for the $\pp\to\ttV$ process in \MGvATNLO{} with the coupling constant \ct in Eq.~\eqref{eq:intro-zprime} set to unity, corresponding to a 4\% width, then scaled to the relevant width using the relation $\GammaV/\mV \approx \ct^{2}/(8\pi)$. The larger relative widths should be regarded as benchmark parameterisations of broad resonant signals. Further details on the model can be found in Refs.~\cite{Greiner:2014qna, Kim:2016plm}.
\par}

Alternative interpretations assuming a scalar or pseudoscalar mediator are performed following an effective field theory approach~\cite{Artoisenet:2013puc}, where a heavy Higgs boson is introduced. Corresponding signal samples $\pp\to\ttS$ and $\pp\to\ttP$ are generated at LO using \MGvATNLO{} v2.6.5 (2016--2018) or v2.9.18 (2022) with the same resonance mass and decay width parameters as used in the simulation of \partV boson production, and again only considering diagrams with the mediator in the $s$-channel.

{\tolerance=600
The production of \ttbar, single top quark, and \ttbar in association with a Higgs boson (\ttH{}) background processes are simulated at next-to-leading order (NLO) in quantum chromodynamics (QCD) using \POWHEG v2.0~\cite{Nason:2004rx,Frixione:2007vw,Alioli:2010xd,Re:2010bp,Frixione:2007nw,Hartanto:2015uka}, while the SM four top quark production (\fourt) process is generated at NLO using \MGvATNLO. 
Drell--Yan production with up to four additional jets (\zjets) with MLM matching~\cite{Alwall:2007fs,Alwall:2008qv}, top quark production in association with gauge bosons and up to one additional jet (\ttZ, \ttW), and three top quark production (\threet) are generated at LO using \MGvATNLO, while diboson processes (\VV with $\PV=\PZ,\PW$) are generated at LO using \PYTHIA. \MGvATNLO{} v2.6.5 is used for 2016--2018, and v2.9.18 for 2022 for all SM processes.
 \par}

Where possible, major backgrounds are normalised to higher-order inclusive cross section calculations. 
The \ttbar cross section is taken from a calculation at next-to-NLO in QCD (NNLO) with next-to-next-to-leading logarithmic soft-gluon resummation using the \textsc{Top++} v2.0 program~\cite{Czakon:2011xx}, while the single top cross sections are from an NNLO calculation using the \textsc{MCFM} program~\cite{Campbell:2020fhf}. The \fourt cross sections include both QCD and electroweak NLO corrections~\cite{Frederix:2017wme,vanBeekveld:2022hty}.

{\tolerance=600
All samples are generated using the NNPDF3.1~\cite{NNPDF:2017mvq} NNLO parton distribution function (PDF) set and parton showering is performed with \PYTHIA{} v8.240 (2016--2018) or v8.306 (2022)~\cite{Sjostrand:2014zea} using the CP5~\cite{Sirunyan:2019dfx} underlying event tune. The effect of pileup is accounted for by overlaying additional simulated minimum bias collisions, matched to the distribution in data per data-taking year. All generated events undergo a full \GEANTfour~\cite{geant} simulation of the detector response and are subjected to the same event reconstruction as the data.
 \par}

\section{Event selection and categorisation}
\label{sec:selection}

This analysis targets the production of a \partV boson in association with \ttbar, where both associated top quarks decay in the leptonic channel $\PQt \to \PQb\Pell\PAGn$ (\tlept).
The \partV boson decays into a further \ttbar pair, which both decay hadronically, $\PQt \to \PQb\PQq\PAQqpr$ (\thad).
The associated top quarks result in two leptons $\Pell=\Pe,\PGm$ either directly or via subsequent \Pgt lepton decays, 
two bottom quarks (\PQb), and two neutrinos in the event.
The hadronically decaying top quarks from the resonance are reconstructed from \jetJ jets, as they are Lorentz-boosted and their decay products become collimated.

\subsection{Lepton selection}

Leading (subleading) \pt electron candidates must have $\pt>25$ (15)\GeV and $\abs{\eta}<2.5$. 
A boosted decision tree (BDT) is used to identify genuine electrons with an efficiency of ${\approx}90\%$ as measured in Drell--Yan events~\cite{CMS:2020uim}. 
This BDT is trained using several features of the electron track and of the associated energy deposits in the ECAL, including features sensitive to the electron isolation.

Leading (subleading) \pt muon candidates must be within $\abs{\eta}<2.4$, 
have a \pt of at least 25 (15)\GeV, and fulfil additional identification requirements optimised for the selection of genuine muons that result in efficiencies of ${>}96\%$ as determined from a data sample of Drell--Yan events~\cite{CMS:2018rym}. 
Furthermore, each muon candidate is required to be isolated with a relative isolation parameter ${<}15\%$, defined as the scalar \pt-sum of PF photons and charged and neutral PF hadrons within a cone of radius $\Delta R =0.4$, divided by the muon \pt. 
Contributions from pileup are estimated from the energy deposited by charged hadrons within the isolation cone that are associated with pileup vertices and subtracted from the isolation parameter.

Events with exactly two selected leptons with opposite electric charge are considered for further analysis. This results in event categories $\Pep\Pem$, $\PGmp\PGmm$, and $\Pepm\PGmmp$. Decays of heavy-flavour resonances and contributions from low-mass Drell--Yan processes are vetoed by requiring the invariant mass of the lepton pair (\mll) to be larger than 20\GeV in $\Pep\Pem$ and $\PGmp\PGmm$ events.

\subsection{Fixed-radius jet selection}

Candidate jets from the hadronisation of the two \PQb quarks originating from the \tlept decays are identified using small radius jets (\jetAK) with $\abs{\eta}<2.4$ and $\pt>30\GeV$. Jets that overlap with a selected lepton within $\Delta R<0.4$ are discarded, as are those that overlap with a large \jetJ jet. 
In 2016--2018 data, additional selection criteria are applied to jets with $\pt<50\GeV$ to remove jets consistent with pileup interactions~\cite{CMS-DP-2020-020}.

The \deepjet algorithm~\cite{Bols:2020bkb,CMS:2017wtu} is used to identify jets that originate from the hadronisation of a \PQb quark. 
It is based on a deep neural network and uses detailed information from all constituent PF candidates and reconstructed secondary vertices as input. 
A loose working point is chosen corresponding to an identification efficiency of approximately 90\% for genuine \PQb jets in simulation. 
Misidentification efficiencies for jets originating from \PQu, \PQd, or \PQs quarks or gluons are found to be ${\approx}10\%$, while efficiencies of ${\approx}40\%$ are found for jets from \PQc quarks. 
The efficiencies in simulation are corrected to match the ones observed in data using a combination of measurements based on \ttbar and multijet events, where the latter comprises jets containing muons.

\subsection{Top quark jets with variable radius}
\label{sec:hotvr_jets}

The reconstruction of Lorentz-boosted, hadronically decaying top quarks using large-radius jets with a fixed distance parameter is a well-established technique at the CERN LHC~\cite{Schatzel:2013wsr, Larkoski:2017jix, Asquith:2018igt}. To consistently include all decay products of a top quark with moderate \pt (200--600\GeV) can require a distance parameter as large as $R=1.5$. At higher \pt, however, this approach leads to overly large jets, reducing reconstruction efficiency and increasing susceptibility to contamination from multiple merged light flavour jets~\cite{Kogler:2021kkw}. The ``heavy object tagger with variable R'' (\hotvr) algorithm~\cite{Lapsien:2016zor,CMS:2020poo} addresses this issue by employing a \pt-dependent distance parameter, $R_\text{eff}=600\GeV/\pt$, during sequential recombination with a Cambridge--Aachen distance metric~\cite{Dokshitzer:1997in,Wobisch:1998wt}. Additionally, a mass jump veto~\cite{Stoll:2014hsa} is used to suppress the clustering of PF candidates from soft radiation.

Jets reconstructed with the \hotvr algorithm are required to have a \pt of at least 200\GeV and $\abs{\eta}<2.4$. A BDT has been trained to identify (tag) jets from \thad decays, with separate trainings in each data-taking era. It uses as input the jet mass, the number of subjets as defined by the mass jump condition in the \hotvr clustering, the minimum pairwise subjet mass, the \pt fraction of the leading subjet relative to the jet, and the ratio between 3-subjettiness and 2-subjettiness (which can help distinguish jets with three prongs, as expected for top quarks, from those with fewer prongs)~\cite{Thaler:2010tr}. These inputs are the same as used for \PQt tagging jets by applying selection-based criteria and have been found to be generally well modelled by the simulation~\cite{CMS:2020poo}. The cut-based selection~\cite{CMS:2020poo} defined thresholds in bins of \jetJ jet \pt, all of which have a mistagging rate between 2.6 and 3.0\%. The BDT presented here shows a demonstrable improvement in sensitivity for the same mistagging rate: for $200<\pt<400\GeV$, the BDT has an efficiency of 32\% compared to 20\% for the cut-based approach; for $400<\pt<800\GeV$, 42\% compared to 32\%, and for $\pt>800\GeV$, 57\% compared to 47\%.

For this analysis, a single BDT tagging working point for all values of \jetJ jet \pt is selected based on expected sensitivity. The optimal working point is found to have an efficiency of 84\% and a misidentification rate of 22\%.

\subsection{Event categories}
\label{sec:event_categories}

Events are required to contain exactly two selected leptons and at least two \jetAK jets. The resulting events are further categorised into orthogonal regions based on the number of \PQb-tagged \jetAK jets, \jetJ jets, and \jetJ jets that are additionally \PQt tagged. Two signal regions (SRs) are considered, requiring at least two \PQt-tagged \jetJ jets and exactly one or at least two \PQb-tagged \jetAK jets, which are denoted SR1b2T and SR2b2T, respectively. Additionally, contributions from \zjets production are suppressed by vetoing events within a window of $80<\mll<101\GeV$ around the \PZ boson mass in \Pep{}\Pem{} and \PGmp{}\PGmm{} events. Transfer regions (TRs) are defined, with at least two \jetJ jets and one or two \PQb-tagged \jetAK jets and inverted \PQt tagging requirements as TR1b0T or TR2b0T, respectively. The resulting regions have similar kinematic characteristics to the SRs and hence are used for the background estimation.

In events with at least two \jetJ jets, the invariant mass \mJJ is calculated from the four-momentum sum of the leading-\pt and subleading-\pt \jetJ jets. In signal events, \mJJ approximates the mass of the decaying \partV boson and is therefore a sensitive observable in this analysis since contributions from residual background processes are found to exhibit a falling \mJJ spectrum.

A set of control regions (CRs) are defined for \Pep{}\Pem{} and \PGmp{}\PGmm{} events by inverting the \zjets process veto, \ie selecting events explicitly with $80<\mll<101\GeV$. The resulting events are required to contain at least two \jetJ jets (CR2J) and are further categorised by the number of \PQt-tagged jets into two regions: one labelled CR2J0T with no \PQt-tagged jets among the \jetJ jets, and another labelled CR2J2T where both \jetJ jets are \PQt tagged. These CRs are dominated by \zjets events, where the extra jets are used to validate the selection and analysis strategy---in particular the procedure used to determine the background yield in the SR from data, as described in Section~\ref{sec:background}.

The object requirements of the different regions are summarised in Table~\ref{tab:selections}.

\begin{table*}[!tb]
\centering
\caption{
  Selections for the control, transfer, and signal regions. The number of leptons required is denoted $n_{\Pell}$, $n_{\jetAK}$ refers to the number of \jetAK jets and $n_{\PQb}$ to the number of those that are required to be \PQb tagged. In each region, $n_{\jetJ}$ \jetJ jets are required and $n_{\PQt}$ of these are required to be \PQt tagged.
}
\begin{tabular}{l c c | c c | c c }
& \multicolumn{2}{c|}{CR2J} & \multicolumn{4}{c}{} \\
& CR2J0T & CR2J2T & TR1b0T & TR2b0T & SR1b2T & SR2b2T \\
\hline
$n_{\Pell}$ & \multicolumn{2}{c|}{${=}2$ ($\Pe\Pe$, $\PGm\PGm$)} & \multicolumn{4}{c}{${=}2$ ($\Pe\Pe$, $\Pe\PGm$, $\PGm\PGm$)} \\ 
\mll & \multicolumn{2}{c|}{$\in[80, 101]$\GeV} & \multicolumn{4}{c}{$\notin[80, 101]$\GeV ($\Pe\Pe$, $\PGm\PGm$)} \\ 
$n_{\jetAK}$ & \multicolumn{2}{c|}{${\geq}2$} & \multicolumn{2}{c|}{${\geq}2$} &\multicolumn{2}{c}{${\geq}2$} \\ 
$n_{\PQb}$ & \multicolumn{2}{c|}{\NA} & ${=}1$ & ${\geq}2$ & ${=}1$ & ${\geq}2$ \\  
$n_{\jetJ}$ & \multicolumn{2}{c|}{${\geq}2$} & \multicolumn{2}{c|}{${\geq}2$} & \multicolumn{2}{c}{${\geq}2$} \\ 
$n_{\PQt}$ & ${=}0$ & ${\geq}2$ & ${=}0$ & ${=}0$ & ${\geq}2$ & ${\geq}2$ 
\end{tabular}
\label{tab:selections}
\end{table*}

\subsection{Top quark jet identification performance}

The performance of the BDT classifier score used to \PQt tag \jetJ jets is investigated in a separate study using a different selection and data set compared to the main analysis. 
It utilises a sample enriched in \ttbar events, where one top quark decays hadronically and one decays leptonically to a muon and a \PQb-tagged \jetAK jet. 
Events are selected using a single-muon trigger requiring a muon with a \pt of at least 50\GeV within $\abs{\eta}<2.4$. A region enriched in boosted leptonically decaying \PW bosons from \ttbar decays is defined by selecting events with at least one \PQb-tagged \jetAK jet in the same hemisphere as the muon, and at least one \jetJ jet. The muon must also be separated from the \jetJ jet in $\phi$ by more than two radians, and the absolute value of the sum of the muon \ptvec and \ptvecmiss must exceed 250\GeV. After these selections, approximately 90\% of the events come from \ttbar processes, with smaller contributions from \PW{}+jets, single top quark production, and Drell--Yan processes.

Binned maximum likelihood fits are performed using the jet mass distribution of the leading \jetJ jet in categories passing and failing the BDT cutoff to estimate correction factors of the \PQt tagging efficiencies in simulated events. Three fit templates are constructed from simulated background samples as used in the main analysis by splitting events depending on the \jetJ jet origin as inferred from generator-level information. Events are considered ``\PQt-matched'' if a \PQb quark and two light quarks from a hadronically decaying top quark are found within the \jetJ cone (\ie the distance in the $\eta$-$\phi$ plane from the centre of the \jetJ jet is less than $R_\text{eff}$ of that jet). The background events are then split into two templates, which reflect processes with different kinematic properties: ``partial \PQt-matched'' are events where either two quarks from a \PW boson decay or a \PQb quark and only one quark from a \PW boson decay are found within the \jetJ cone, whilst ``non-matched'' events comprise all other \jetJ jets.

\begin{figure}[htbp]
	\centering
	\includegraphics[width=0.95\linewidth]{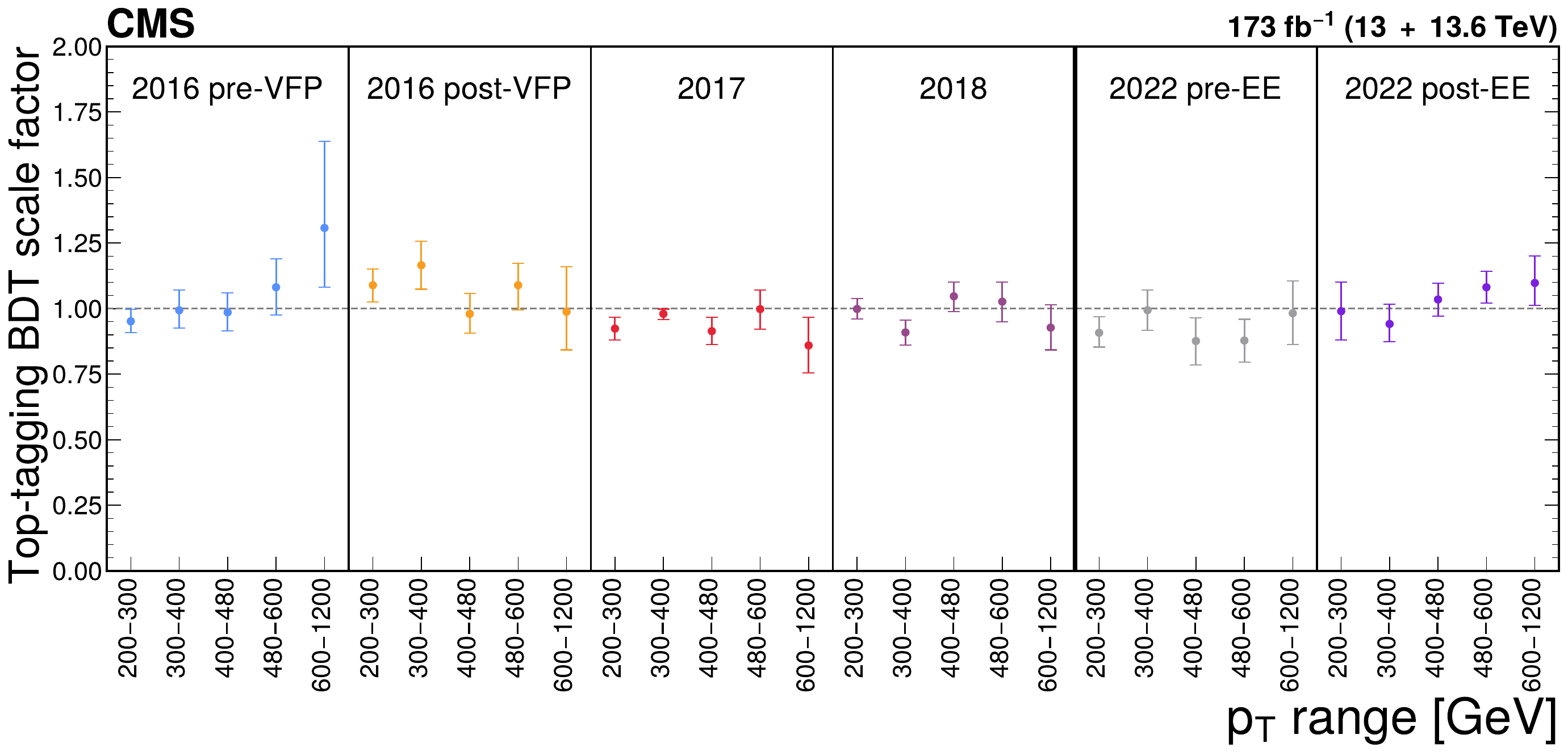}
    \caption{Data-to-simulation scale factors for the BDT classifier to identify \PQt jets in bins of \jetJ jet \pt for each data-taking period.}
	\label{fig:selection-bdt-sfs}
\end{figure}

Separate fits are performed for each data-taking period in bins of \jetJ jet \pt, with edges at 200, 300, 480, 600, and 1200\GeV. Various sources of systematic uncertainties, as described in Section~\ref{sec:sys}, are incorporated into the fit through nuisance parameters. In the main analysis, the resulting shifts in the yields of the ``\PQt-matched'' template per \pt bin are applied as correction factors to the \PQt tagging efficiencies in simulated events. These factors are shown in Fig.~\ref{fig:selection-bdt-sfs}. The dominant source of uncertainty is the statistical uncertainty in the data.

\begin{figure*}[tbp]
	\centering
	\includegraphics[width=0.49\textwidth]{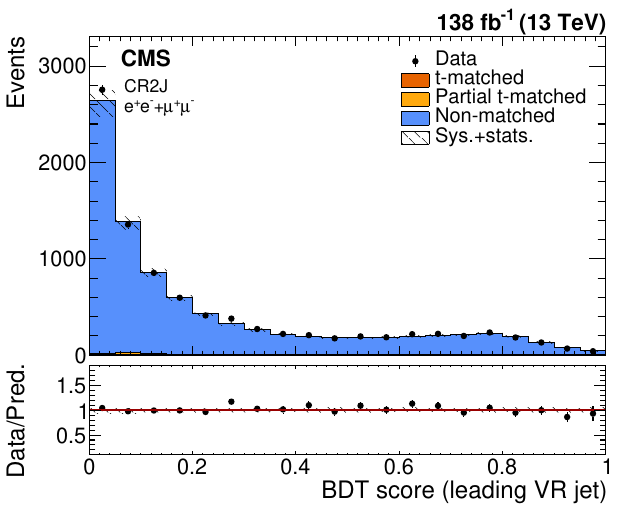}
	\includegraphics[width=0.49\textwidth]{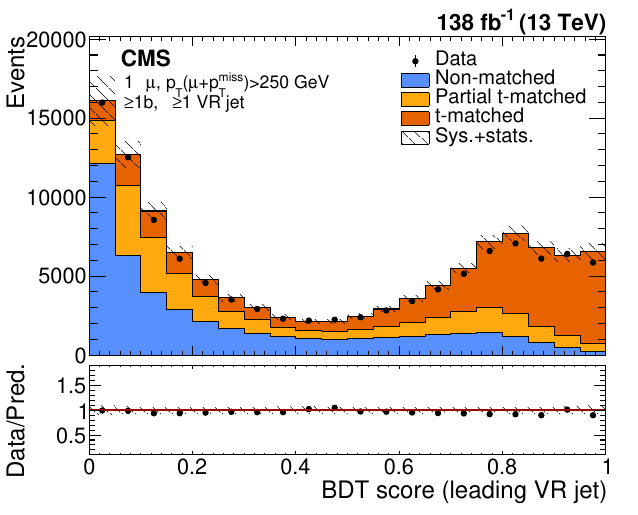}\\
	\includegraphics[width=0.49\textwidth]{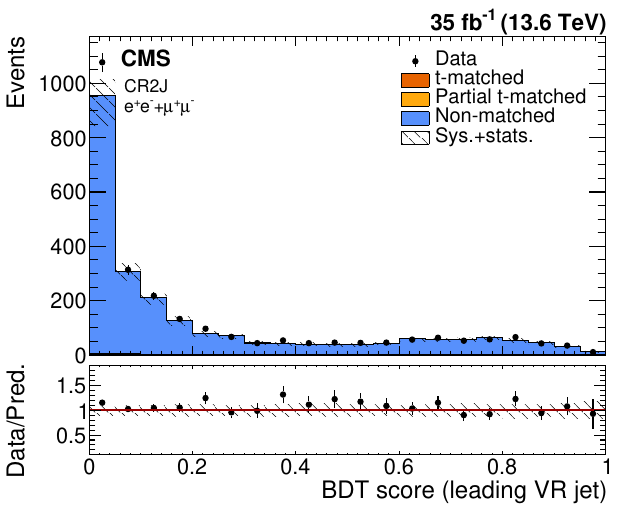}
	\includegraphics[width=0.49\textwidth]{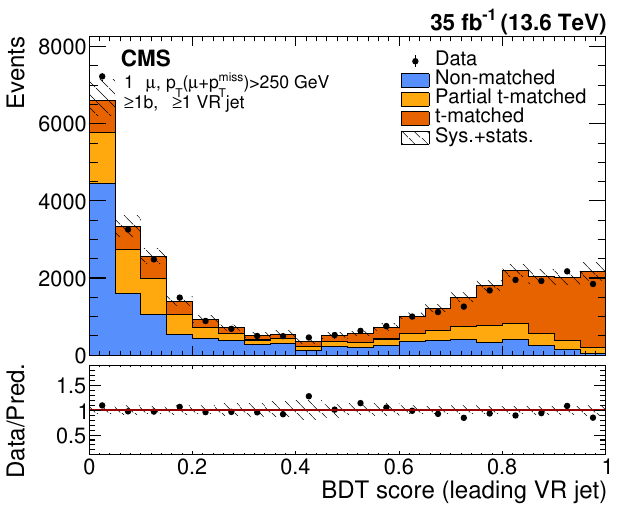}
	\caption{\label{fig:selection-bdt}Distributions of the BDT classifier score in regions enriched in \zjets{} (left) and \ttbar (right) events for the 2016--2018 data-taking periods (upper row) and the 2022 data-taking period (lower row). The simulated backgrounds are split by the origin of the leading \jetJ jet.}
\end{figure*}

Figure~\ref{fig:selection-bdt} shows the distribution of the BDT classifier score after application of the scale factors in the \ttbar-enriched region of this study, and for comparison also in CR2J of the main analysis, where the latter is depleted of genuine top quark jets and hence allows an evaluation of the performance at rejecting background jets. No dedicated misidentification factors are computed since the background from mistagged \PQt jets in the final fit is estimated from data in the TRs. Nevertheless, the data are found to be well modelled in both regions.

\section{Background estimation}
\label{sec:background}

{\tolerance=600
Processes with multiple jets constitute the main source of background in the SRs. The production of \ttbar with additional jets accounts for 85 and 75\% of the total expected background in SR2b2T and SR1b2T, respectively, as estimated from simulation. The majority of this background comes from events with correctly identified leptons and \PQb jets, with additional light jets misidentified as \PQt jets, while in some cases the \PQb jets can be misidentified as \PQt jets, with other light jets misidentified as \PQb jets. A smaller fraction, about 4\% of the total, can be attributed to events containing genuine hadronically decaying top quarks that are primarily produced in $\ttbar\PV$, \ttH, and processes with three or four top quarks, while a further 1--3\% can be attributed to other heavy objects (\PZ and Higgs bosons, as well as \PW bosons not from top-quark decays). A summary of the origins of \jetJ jets in the TRs and SRs is provided in Table~\ref{tab:jet_origs}.
\par}

\begin{table}[htbp]
   \centering
   \caption{Background jet origins (in \%) in each signal or transfer region, as predicted by simulation. Jet origins are determined by matching \jetJ jets geometrically to generator-level particles, requiring the distance in the $\eta$-$\phi$ plane to be less than $R_\text{eff}$.}
   \cmsTable{
       \begin{tabular}{@{}lcccc@{}}
           Jet origins [\%]                      & TR2b0T & SR2b2T & TR1b0T & SR1b2T \\\hline
           \PQt-matched                          & ${<} 1$ & 4 & ${<} 1$ & 3 \\
           Partial \PQt-matched                  & 1 & 2 & 2 & 3 \\
           Light quarks/gluon                    & 91 & 72 & 84 & 73 \\
           \PQb quark                            & 7 & 21 & 12 & 18 \\
           Non-top-quark heavy objects           & 1 & 1 & 2 & 3 \\
       \end{tabular}
   }
   \label{tab:jet_origs}
\end{table}

We estimate the \mJJ distribution from background processes in the SRs using data in the TRs.
Since the majority of background events originate from misidentified \PQt-tagged jets, the procedure uses reweighted data events with no \PQt-tagged jets. A corresponding transfer factor is calculated in bins of \mJJ as the ratio of the number of events in SR2b2T to TR2b0T and SR1b2T to TR1b0T, respectively, using simulated background samples. The \mJJ binning is optimised separately for the 13 and 13.6\TeV data sets based on the amount of data in the TRs to ensure each bin contains background estimation data in each \PQb jet and lepton category. The starting point for this is a 100\GeV binning, with bins then merged until each contains some data events in each of TR2b0T and TR1b0T. In a few cases, an intermediate bin edge was selected where the original 100\GeV binning would result in a particularly large bin close to the peak of the distribution.

The background estimation from data is only designed to estimate the contribution from misidentified top quarks; events with at least one \jetJ jet matched to a hadronically decaying top quark are estimated directly from simulation. To ensure these events are not also estimated by the data-driven method, events containing genuine hadronically decaying top quarks are first excluded from the simulated events used to derive the transfer factor. Then the contribution from these events in the TRs, estimated from simulation, is subtracted from data in the TRs prior to applying the transfer factor.

\begin{figure*}[tbp]
	\centering
	\includegraphics[width=0.49\textwidth]{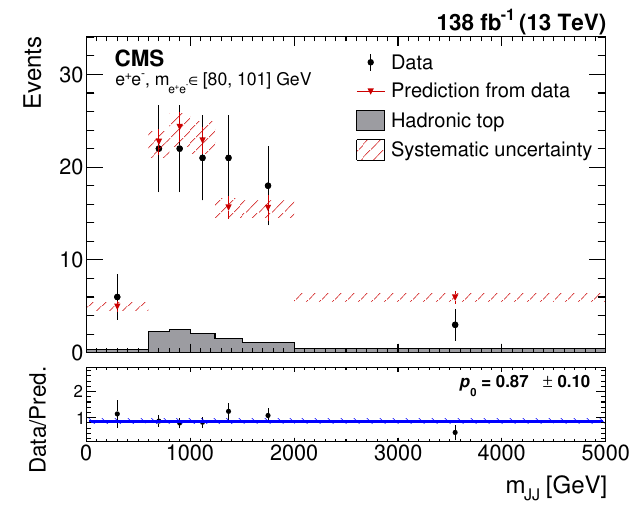}
	\includegraphics[width=0.49\textwidth]{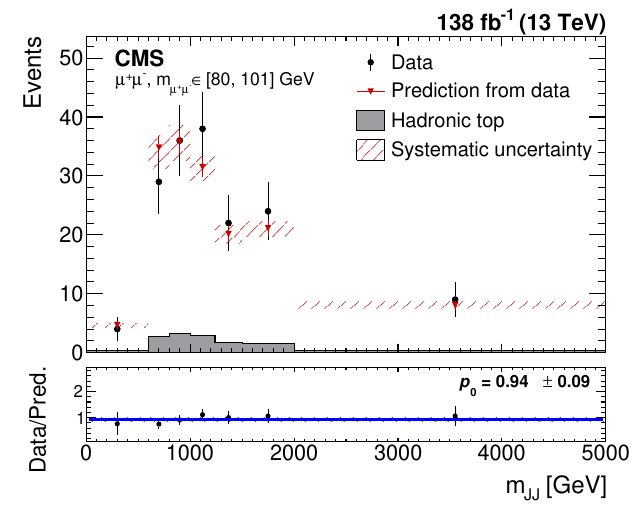}\\
	\includegraphics[width=0.49\textwidth]{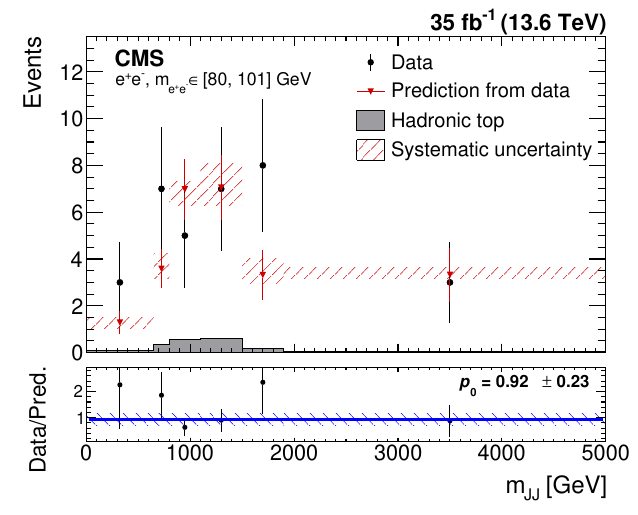}
	\includegraphics[width=0.49\textwidth]{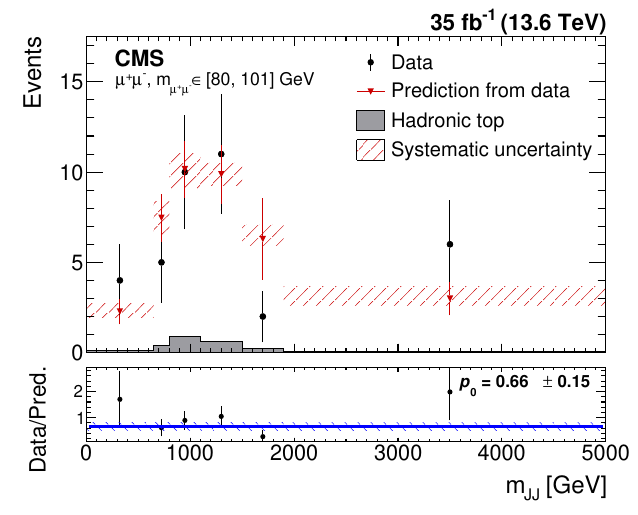}
	\caption{\label{fig:validation} Observed and estimated \mJJ distributions in CR2J2T for 
   \Pep{}\Pem{} (left) and \PGmp{}\PGmm{} (right) events for the 2016--2018 data-taking periods (upper row) and the 2022 data-taking period (lower row). The points for the background prediction (red) include the contribution from hadronic top-quark processes estimated directly from simulation (filled grey bins). The error bars on these indicate the statistical uncertainty from the data in the CR2J0T region, while the error bands indicate the systematic uncertainties from the simulation-based transfer factors (and from the hadronic top contribution). The ratio of the predicted background to the data is shown in the bottom panel, with the ratio determined from a fit using a constant function shown as a blue line.}
\end{figure*}

The method is validated in the \Pep{}\Pem{} and \PGmp{}\PGmm{} CRs defined in Section~\ref{sec:event_categories}. 
The \mJJ distribution of events with two \PQt-tagged jets (CR2J2T) is predicted from events with no \PQt-tagged jets (CR2J0T) using a transfer factor derived in the same way as in the SR. 
Figure~\ref{fig:validation} shows the \mJJ distributions of data in CR2J2T (black round markers), which are compared to the estimated distributions from the transfer factor (red triangle markers). 
The uncertainties in the estimation include contributions from the statistical uncertainty in the data in CR2J0T and the statistical and systematic uncertainties in the simulation used to derive the transfer factor, as documented in Section~\ref{sec:sys}. 
The shapes of the estimated distributions are consistent with the data in all of the CR2J2T regions. However, differences in the total yield between data and the estimated background are found. The efficiencies of \PQt tagging individual \jetJ jets in data and simulation in this region are compared differentially in jet mass and \pt and found to be statistically consistent; the normalisation difference is consistent with a mismodelling of the correlation between the tagging probabilities of the two \jetJ jets in the simulation from which the transfer factors are derived. A constant function is fit to the ratios between the data and the estimated background yields, as shown by the horizontal lines in the ratio panels of Fig.~\ref{fig:validation}. The fits provide a good description of the data in all cases, with $\chi^2/\mathrm{ndf}\lesssim 1$. For the final results, these differences are applied as correction factors on the estimated background yields in the SRs for \Pep{}\Pem{} and \PGmp{}\PGmm{} events, while the geometric mean of both is taken for \Pepm{}\PGmmp{} events. The applicability of these correction factors to the SRs assumes that the residual mismodelling of the correlation between the two \PQt-tagging probabilities is sufficiently similar in the CRs and SRs. Differences arising from the larger fraction of \jetJ jets matched to \PQb quarks in the SRs are covered by a dedicated systematic uncertainty, as described in Section~\ref{sec:sys}.

\section{Systematic uncertainties}
\label{sec:sys}

The expected numbers of signal and background events can be affected by several sources of systematic uncertainties. 
These uncertainties are incorporated into the analysis through nuisance parameters that change the expected event yields and are correlated across all \mJJ bins in the SRs and data-taking periods, unless otherwise noted. 
Uncertainties affecting the simulation are applied not only to the signal and simulated backgrounds, but also propagated to the background estimation from the TRs via their effect on the transfer factors (Section~\ref{sec:background}).

Systematic uncertainties in the lepton reconstruction, identification, and trigger efficiencies are accounted for by varying \pt- and $\eta$-dependent correction factors, derived from data-to-simulation ratios, within their uncertainties. A further systematic uncertainty is assigned to corrections for a minor reduction in L1 trigger efficiency due to mistiming in the forward regions of the ECAL in 2016 and 2017 data-taking periods~\cite{CMS:2020cmk}.

{\tolerance=600
The \jetAK jet energy scale in simulated events is modified for all jets using the uncertainty associated with the \pt and $\eta$-dependent correction factors. The \jetAK jet energy resolution is modified in simulated events by either increasing or decreasing the difference in \pt with respect to the matched jet at the generator level. Random smearing is used when no generator-level jet is found within $\Delta R<0.4$. These variations are also extrapolated in a correlated manner to \jetJ jets by recalculating their four-momenta after applying the variations to all subjets within each \jetJ jet. Additional sources of uncertainties are evaluated for \PQb-tagged jets by varying the corresponding estimated (mis-)tagging probabilities within uncertainties depending on the jet flavour obtained from generator-level information~\cite{CMS:2017wtu}. The uncertainty in the efficiency of tagging \jetJ \PQt jets is handled similarly, though a mistagging uncertainty is not estimated since this is included in the uncertainties in the background estimated from data. Since the fit used to derive these corrections is largely dominated by statistical uncertainties, the \jetJ \PQt jet tagging uncertainties are taken as fully uncorrelated with other uncertainties in this fit.
\par}

An uncertainty of 4.6\% in the total inelastic cross section~\cite{CMS:2018mlc} is used when calculating the expected distribution of the number of pileup interactions in data. This shift is propagated to the simulated samples by reweighting events according to the modified distribution of pileup interactions. The integrated luminosity for the 2016--2018 data-taking period has an overall uncertainty of 0.73\%~\cite{CMS-LUM-17-003,CMS-PAS-LUM-20-001}, while the uncertainty in the integrated luminosity of data recorded in 2022 is 1.4\%~\cite{CMS-PAS-LUM-22-001}.

Uncertainties from missing higher-order corrections to the matrix element calculations of the simulated samples are estimated by varying the factorisation and renormalisation scales by factors of 2 and 0.5, and taking the largest of the independent and correlated variations; these variations are propagated to the analysis by way of additional event weights~\cite{Kalogeropoulos:2018cke}. Only shape and acceptance effects from the selection, which can be up to 5\%, are considered as uncertainties in the fit, while the uncertainties in the background cross sections are taken from higher-order calculations, and those in the signal normalisation are considered part of the theoretical uncertainty instead. Uncertainties from the PDF choice are determined by reweighting each event according to the prescribed variations of the corresponding PDF set, which include a variation of the scale used to evaluate the strong coupling constant, in accordance with the PDF4LHC recommendations~\cite{Butterworth:2015oua}, and generally represent variations of less than 2\%. Additionally, the scales used to evaluate the strong coupling constant in the parton shower simulation are varied independently for the initial- and final-state radiation by factors of 2 and 0.5, which lead to variations up to 10\% in the acceptance of events in the SRs since these affect the momentum and substructure of the \jetJ jets, respectively.

The uncertainties in the background estimation arise from four categories of sources. First are the statistical uncertainties from the TRs (as well as a smaller contribution from the statistical uncertainties due to the finite size of the simulated samples in both TRs and SRs), which are considered separately for each bin of the final template and are treated using the Barlow--Beeston approach~\cite{Barlow:1993dm}.
Second, the modelling uncertainties described in the previous paragraphs are incorporated in the calculation of the transfer factors, and in the subtraction of events with genuine \PQt quark jets from the TR distribution. These are generally relatively small as most systematic effects largely cancel in the ratio.
Third is the statistical uncertainty in the correction factor derived in the CRs to correct the normalisation of the prediction to the data.
The final uncertainty accounts for the differences in jet flavour composition between the CRs and SRs (the SRs have a significantly higher fraction of \jetJ jets from \PQb quarks), which could lead to differences in the normalisation relative to the correction factors derived in the CRs.
A conservative estimate of the size of this effect is made by estimating the number of events with at least one \jetJ jet matched to a \PQb quark from simulation, subtracting this contribution from data in the TRs and excluding these events from the transfer factor computation. This new estimate did not give a significantly different shape of the invariant mass spectrum compared to the nominal estimate; however, it is found to predict about 20\% fewer background events. This difference to the nominal background prediction is taken as a normalisation uncertainty.

The total uncertainty is found to be dominated by the statistical uncertainty in the data in the SRs, which is 10\% larger than the combined systematic uncertainty. The largest systematic uncertainty is attributed to the background estimation, with the main sources being the statistical uncertainty in the data in the TRs and the uncertainty accounting for differences in the jet flavour composition.

\section{Results}
\label{sec:results}

The search is conducted by performing a simultaneous maximum likelihood fit to the data, using templates of the \mJJ distribution for the expected signal and the estimated background as described in Section~\ref{sec:background}.
The fit is performed simultaneously in the three dilepton channels of the SR2b2T and SR1b2T regions at the two centre-of-mass energies.

\begin{figure*}[tbp]
	\centering
	\includegraphics[width=\textwidth]{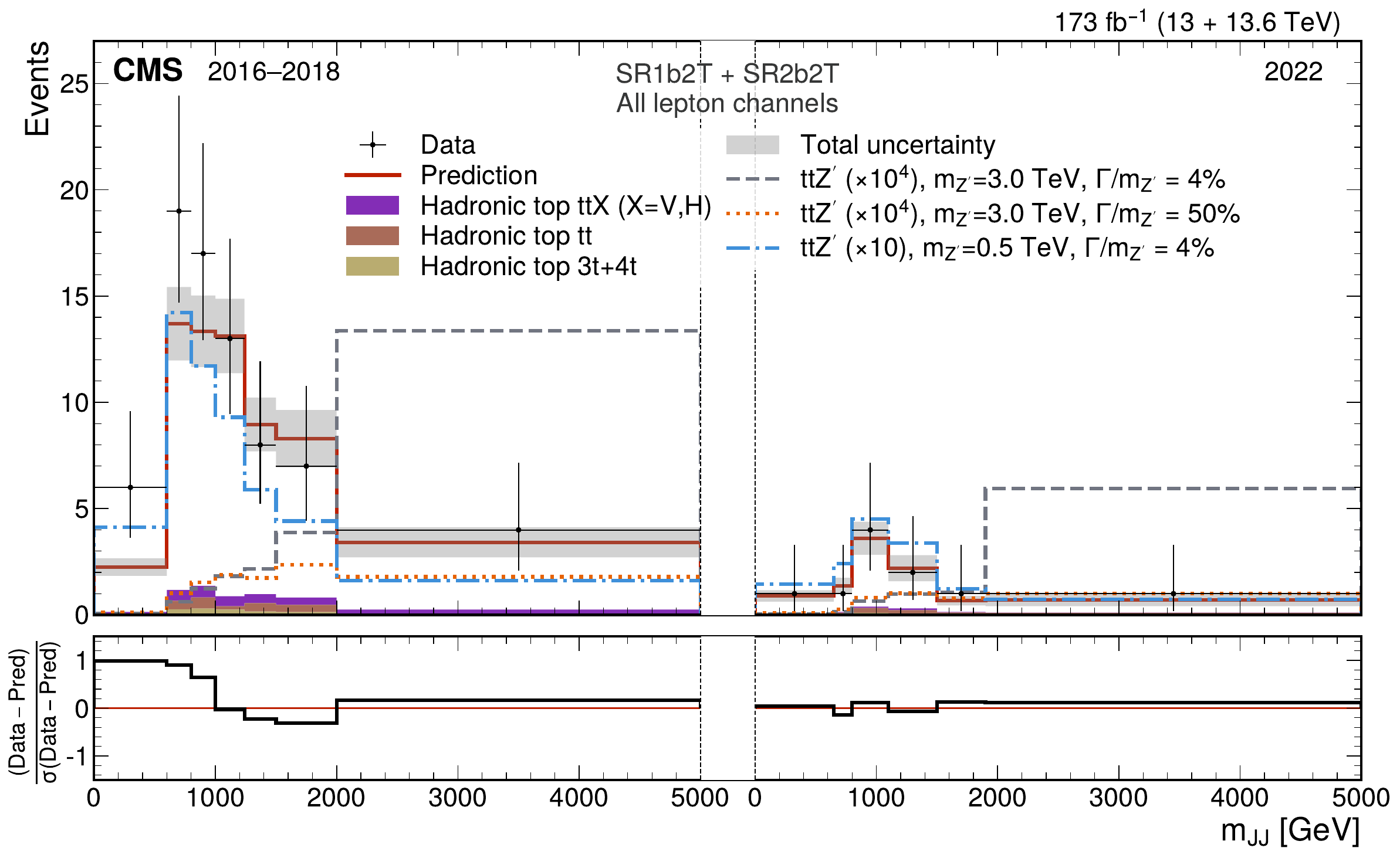}
	\caption{\label{fig:postfitdistributions}The background-only post-fit distribution of \mJJ for the background prediction (red line), including contributions estimated directly from simulation (filled bins) in the merged SR1b2T and SR2b2T for the 2016--2018 (left) and 2022 (right) data-taking periods. The \partV signal scenarios with $\mV=500\GeV$ and $\GammaV/\mV=4\%$, and $\mV=3000\GeV$ and $\GammaV/\mV=4$ and 50\%, are overlaid. The signal cross sections are scaled as indicated in the legend. The pulls (difference of data and prediction, divided by the uncertainty on this difference) are shown in the lower panel.}
\end{figure*}

{\tolerance=600
The systematic uncertainties described in the previous section are incorporated as nuisance parameters, using log-normal probability distributions for normalisation uncertainties and Gaussian distributions for shape-related uncertainties. 
The background-only fit results are shown in Fig.~\ref{fig:postfitdistributions} for the SR2b2T and SR1b2T regions, and the three dilepton channels, merged together, separately for the 2016--2018 (13\TeV) and 2022 (13.6\TeV) data-taking periods. The data agree with the SM background expectation within the uncertainties, and the $p$-value for the background-only fit is 0.21.
\par}

Upper limits on the \partV production cross sections at 95\% confidence level (\CL) are determined as functions of \mV and the relative decay width. 
A modified frequentist approach is employed, which uses the profile likelihood ratio as the test statistic~\cite{CMS-NOTE-2011-005}, the \CLs criterion~\cite{junk, read}, and the asymptotic formulae~\cite{Cowan:2010js} to approximate the distributions of the test statistic under the background-only and signal-plus-background hypotheses. 
The CMS statistical analysis and combination tool \textsc{Combine} is used to derive the results~\cite{CMS:2024onh}.

\begin{figure*}[tbp]
	\centering
	\includegraphics[width=0.42\textwidth]{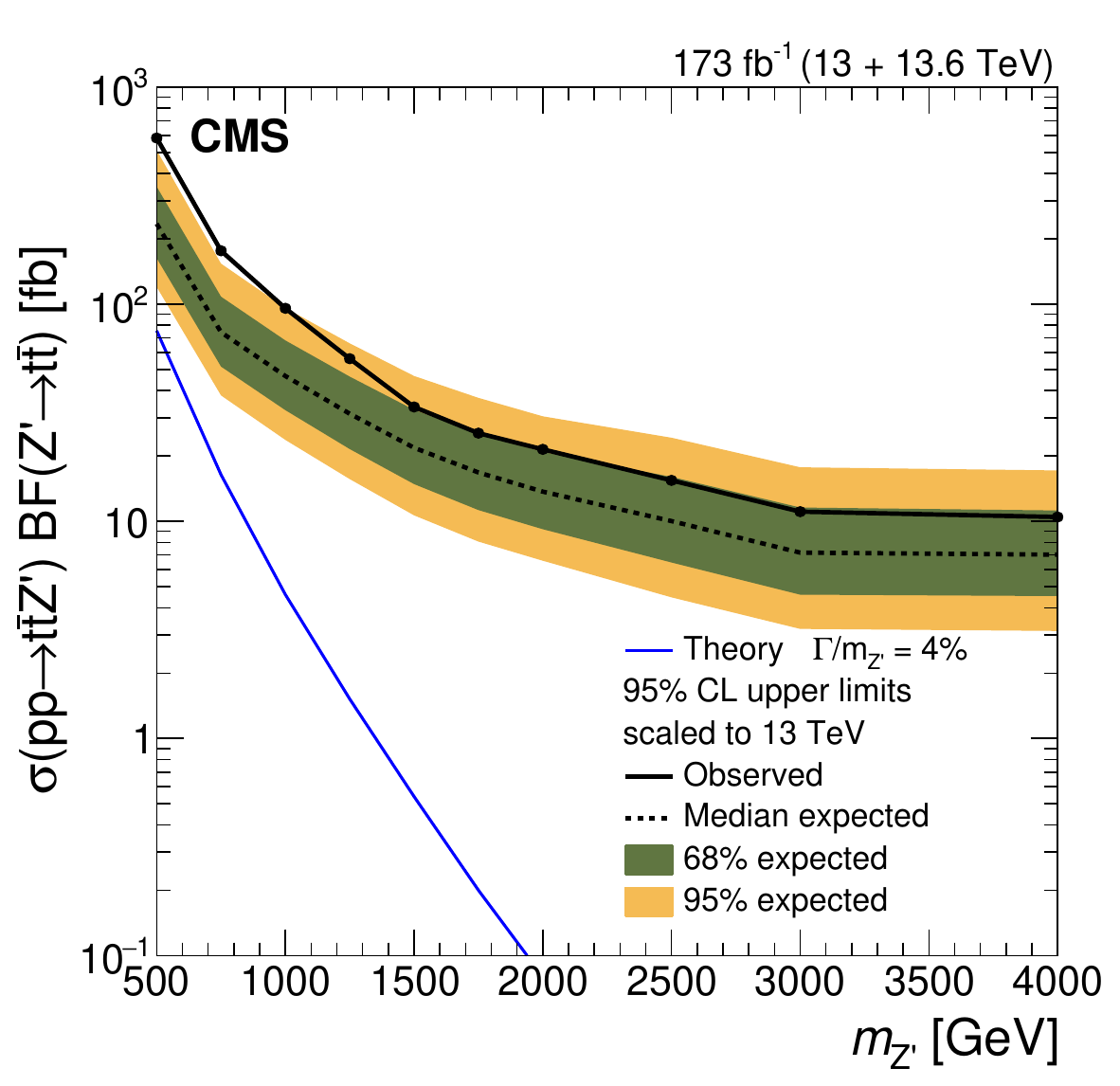}\hspace{0.03\textwidth}
	\includegraphics[width=0.42\textwidth]{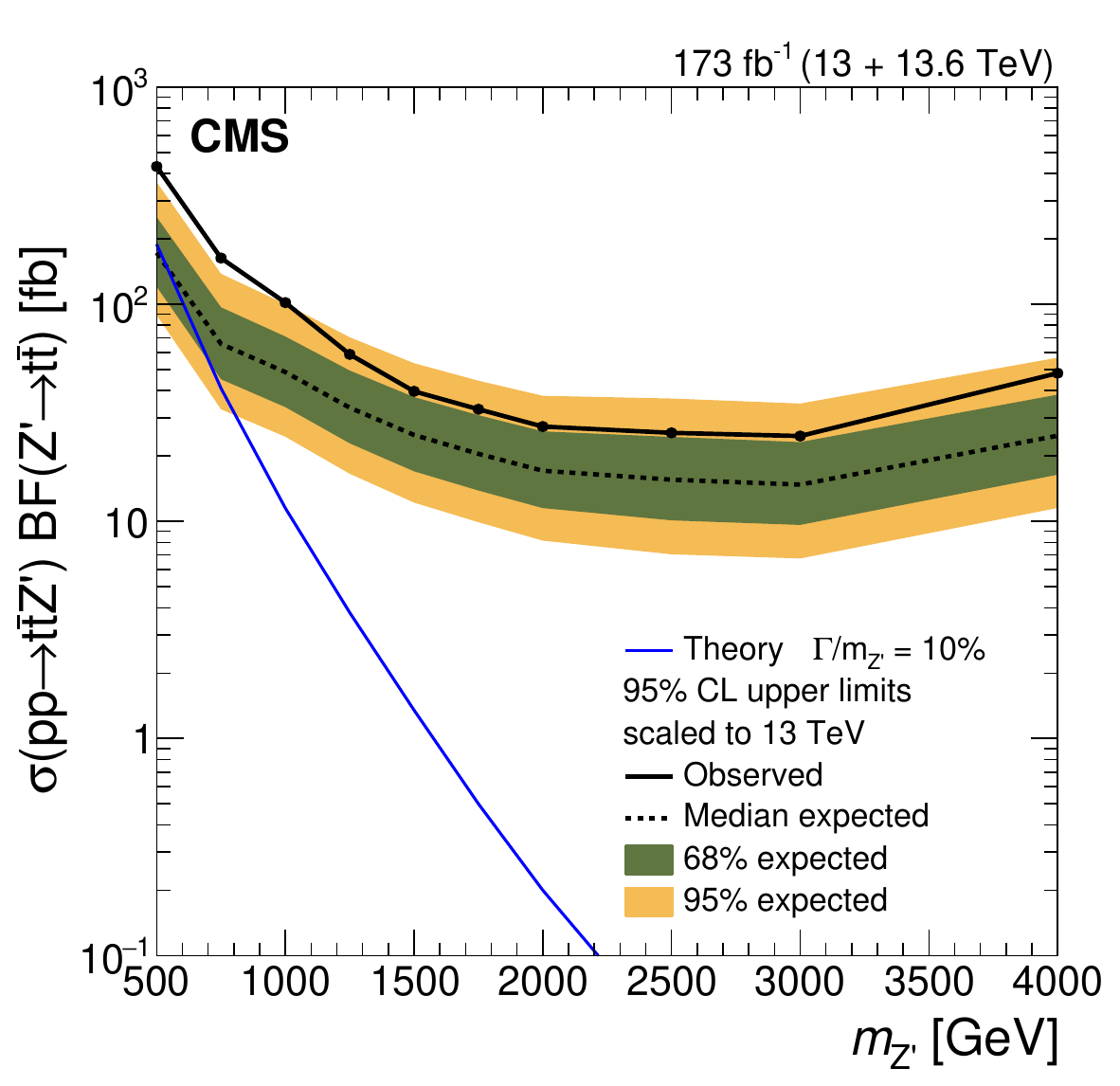}\\
	\includegraphics[width=0.42\textwidth]{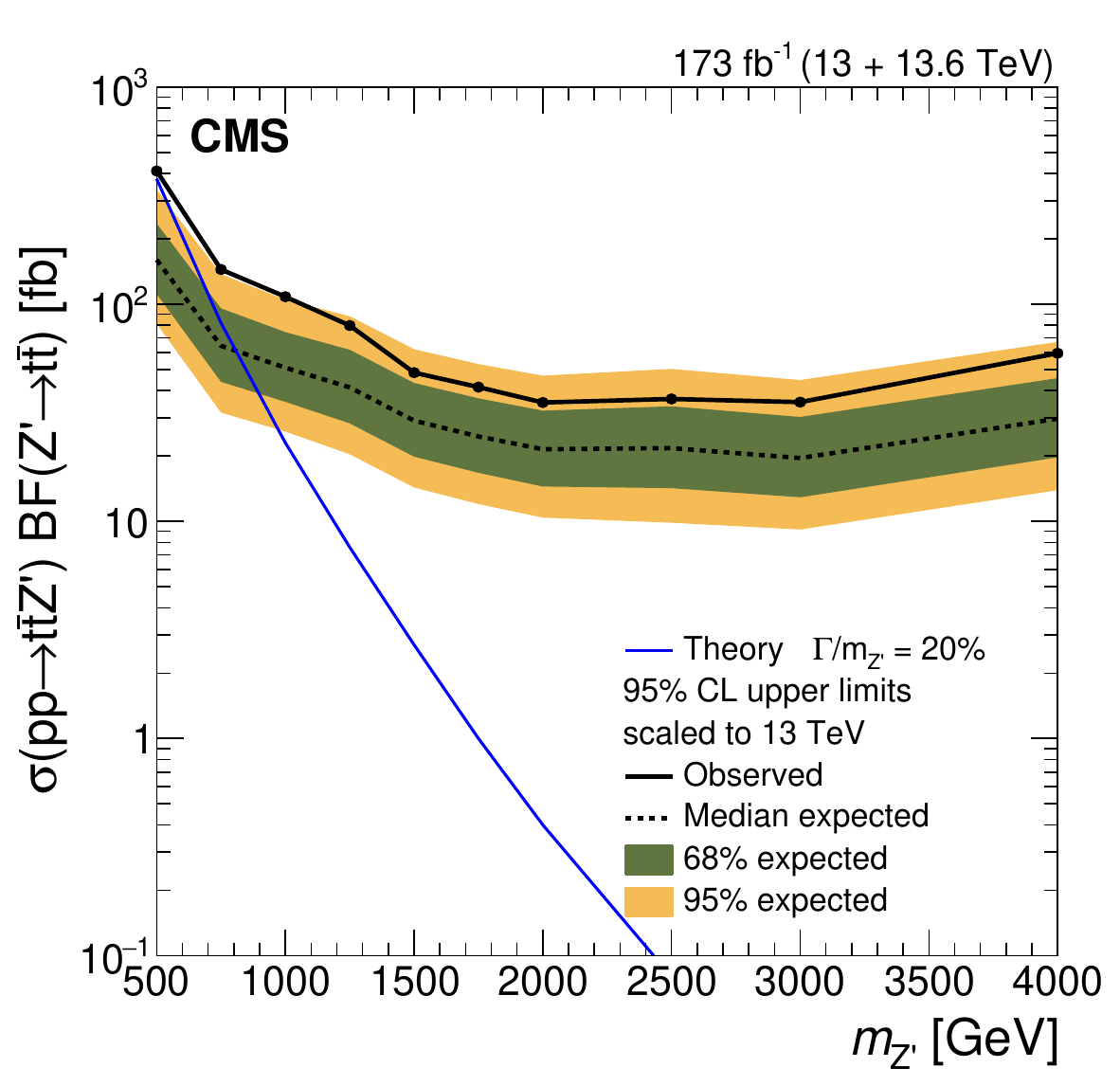}\hspace{0.03\textwidth}
	\includegraphics[width=0.42\textwidth]{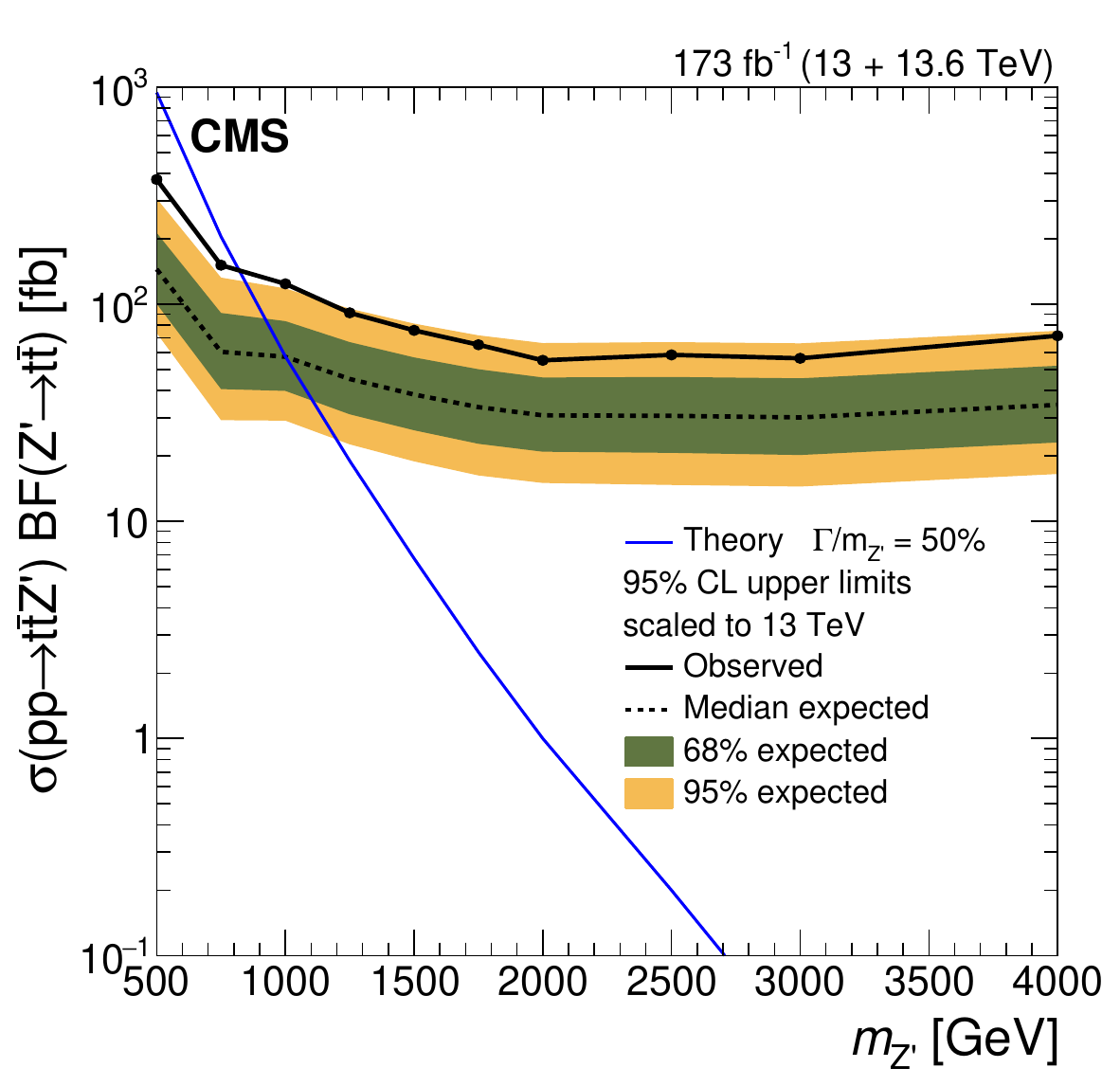}
	\caption{\label{fig:result-limitzprime}Expected and observed upper limits at 95\% \CL on the product of the \ttV production cross section and branching fraction at 13\TeV as a function of \mV for various relative decay widths $\GammaV/\mV$ as indicated in the legends. Data collected at 13.6\TeV are included as described in the text. The theoretical cross section for this model, as described in Section~\ref{sec:samples}, is shown as a solid blue line.}
\end{figure*}
Figure~\ref{fig:result-limitzprime} shows the resulting upper limits for the combination of all data-taking periods and SRs as functions of \mV. Since two different collision energies are combined, the limits are first calculated relative to the theoretical cross sections at the respective energies, then scaled to the \ttV production cross sections at 13\TeV to allow easier comparison with previous results at this energy. The LO theoretical prediction of the 
$\sigma(\pp\to\ttV)$ cross section at 13\TeV, as described in Section~\ref{sec:samples}, is overlaid.
The theoretical prediction includes only the resonant contribution, as stated in Section~\ref{sec:samples}. The resulting exclusions therefore apply to this benchmark signal definition and should not be interpreted as exclusions of a complete top-philic model without accounting for these additional contributions.
For this reference model, this analysis is expected to exclude mediator masses up to 550\GeV for mediators with $\GammaV/\mV=10\%$, 850\GeV for 20\%, and 1000\GeV for 50\%.

Since more events are observed in data than predicted (in total 84 events observed compared to 61.5 expected, pre-fit), particularly at lower \mJJ masses in 13\TeV data, the observed limits are weaker than expected.
This is also observed for the higher mediator masses targeted in this search, which can be attributed to a combination of a slight upwards fluctuation in the final bin, and the broad \mJJ peaks of the signal templates (particularly for larger widths).
Therefore no mediators with 10 or 20\% widths are excluded, and mediators with 50\% relative width are only excluded up to 850\GeV.
The deviation from the background-only hypothesis is larger for lower mediator masses, being largest for the 500\GeV signal with 4\% relative width, with a local significance of 2.1 standard deviations. The best-fit cross section for this signal hypothesis is $340 \substack{+110 \\ -100}\mathrm{(stat.)}\substack{+90 \\ -80}\mathrm{(syst.)}$\unit{fb}.

\begin{figure*}[tb]
	\centering
	\includegraphics[width=0.42\textwidth]{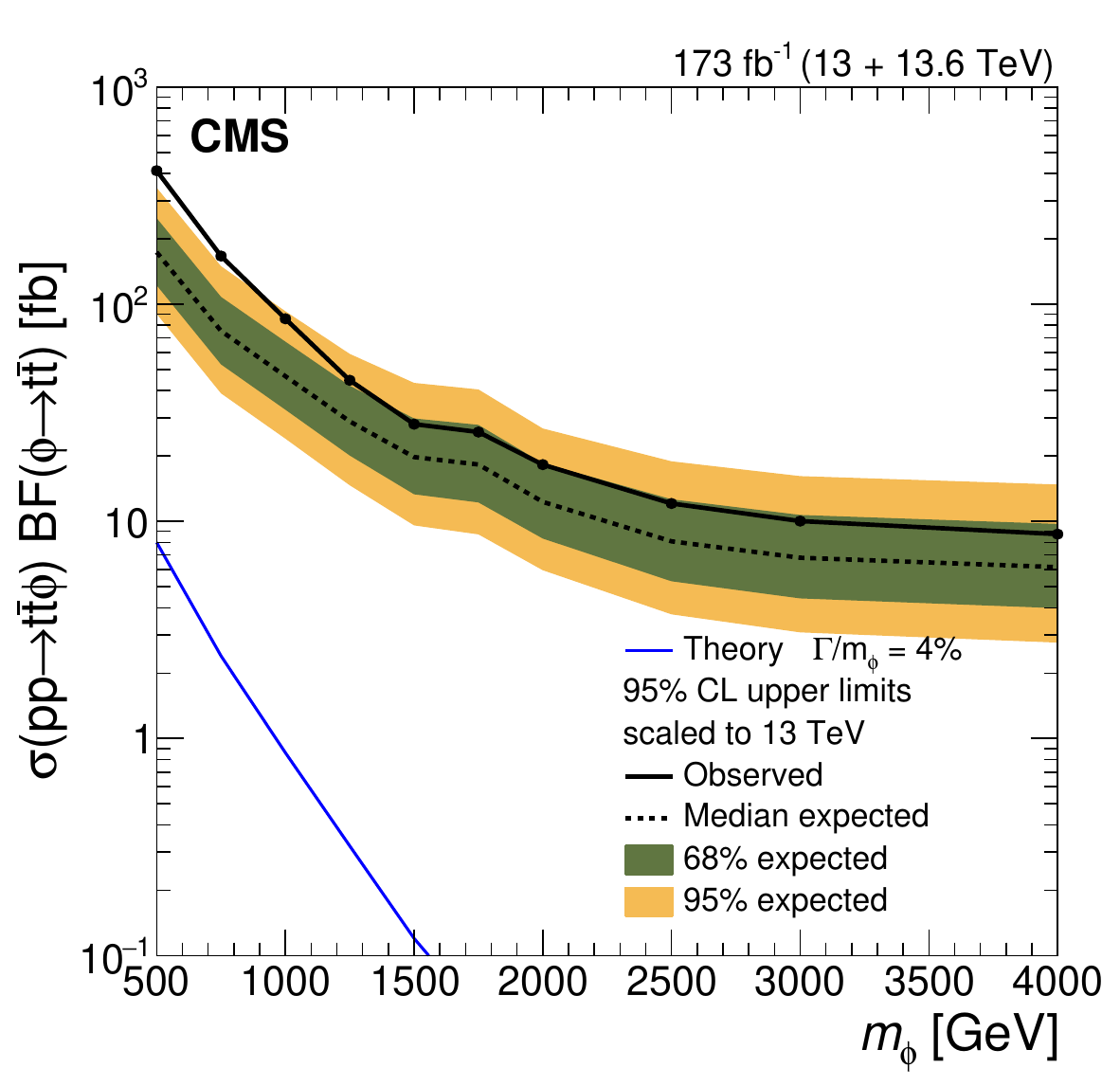}\hspace{0.03\textwidth}
	\includegraphics[width=0.42\textwidth]{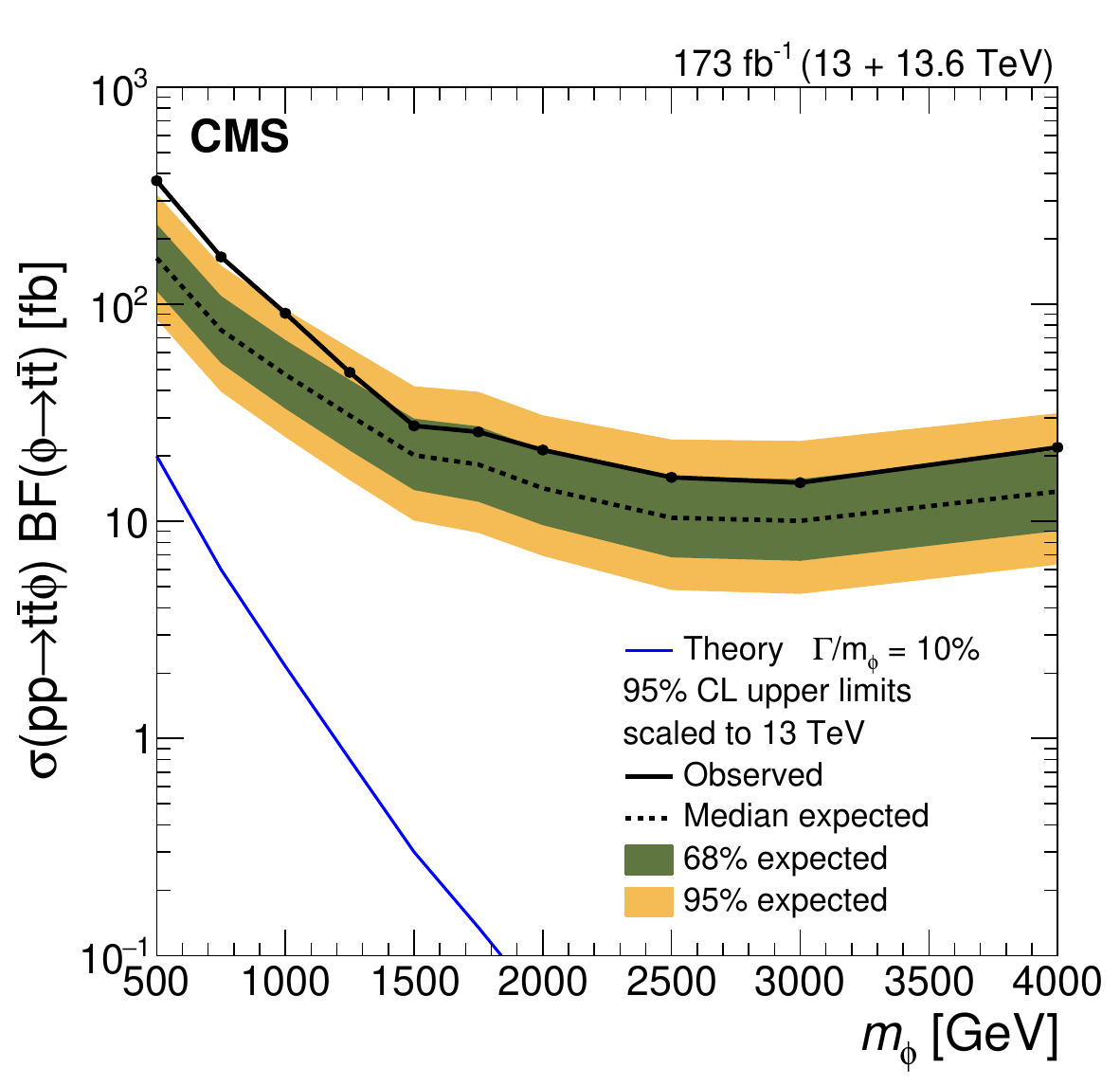}\\
	\includegraphics[width=0.42\textwidth]{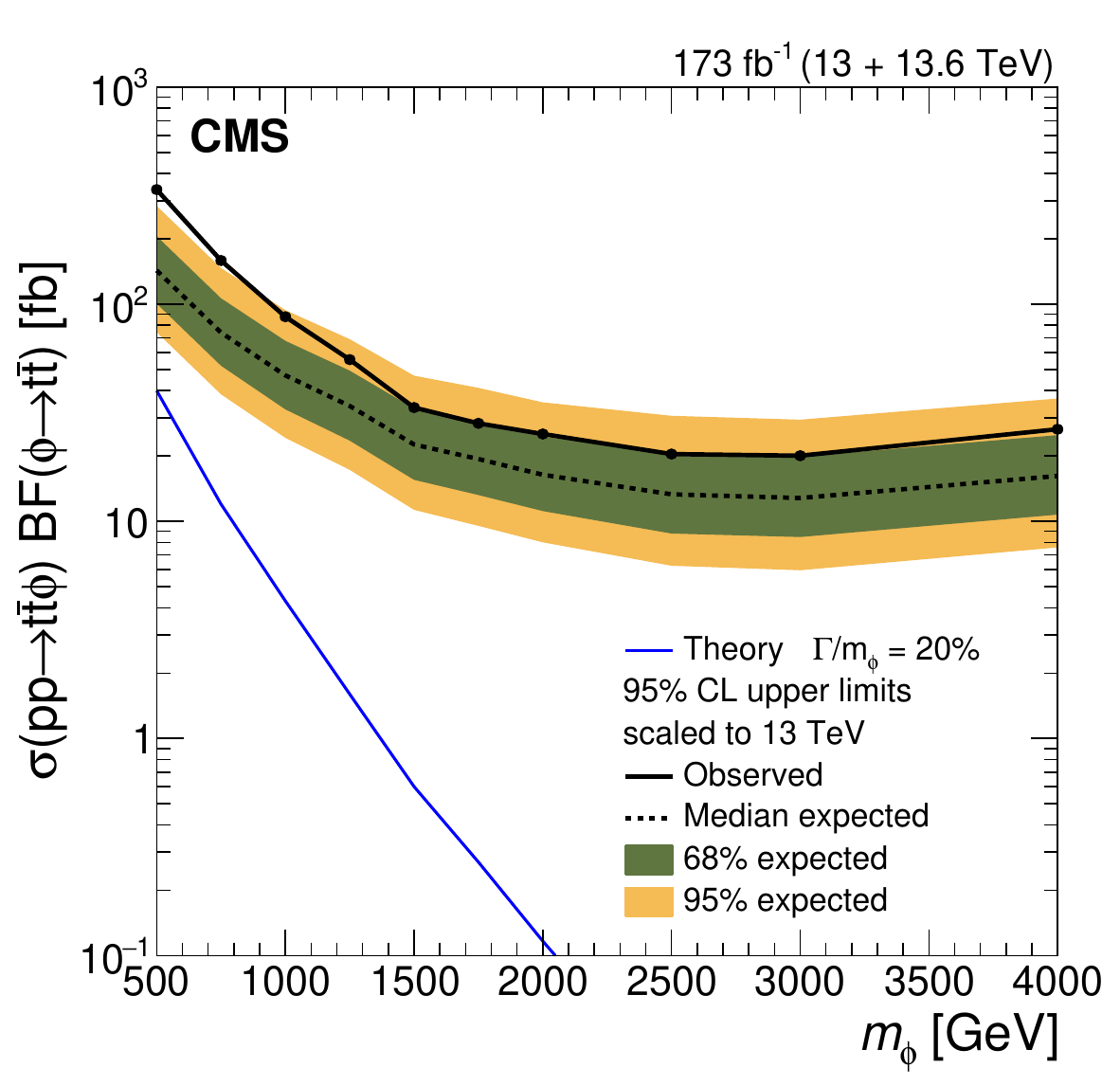}\hspace{0.03\textwidth}
	\includegraphics[width=0.42\textwidth]{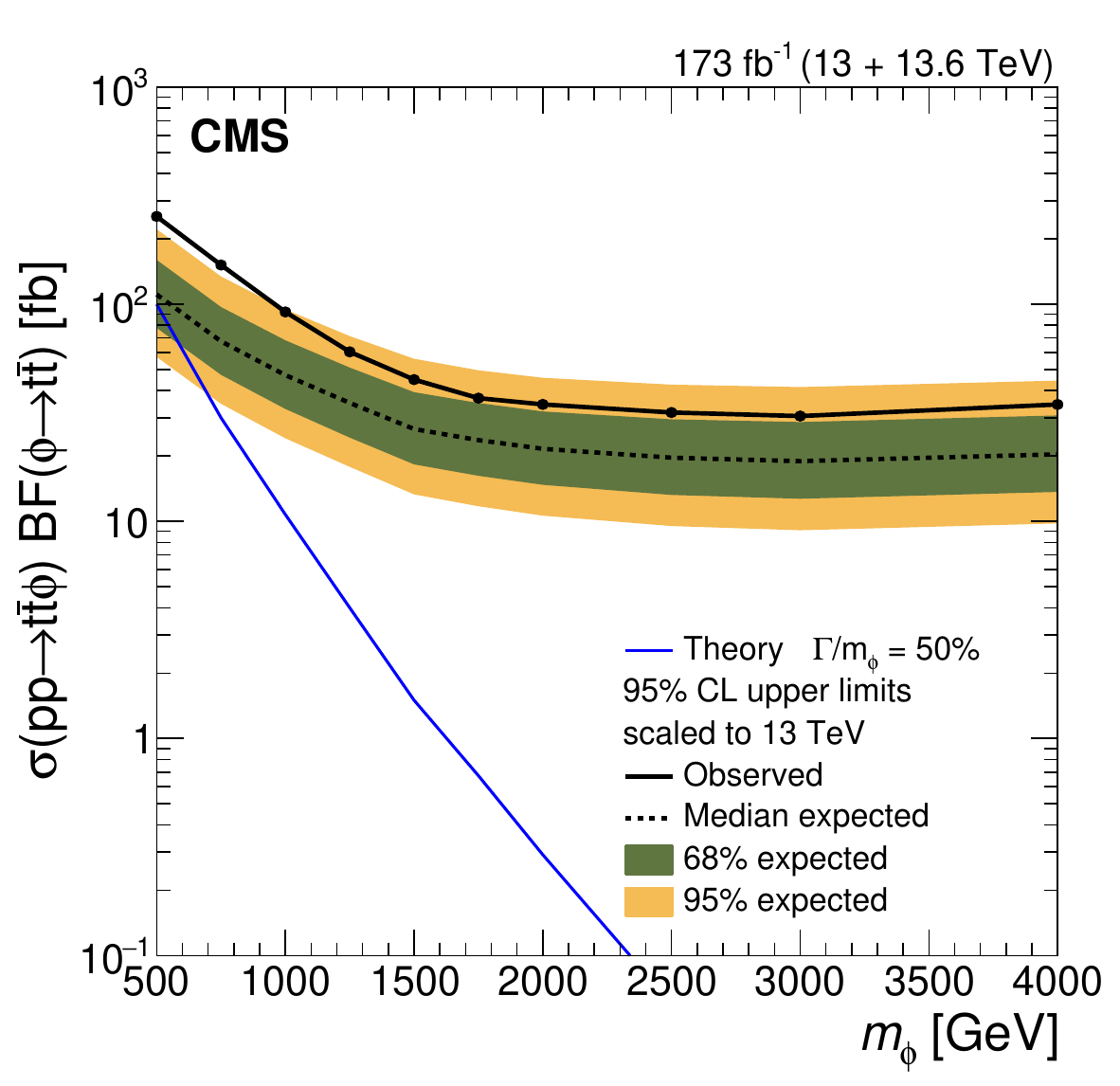}
	\caption{\label{fig:result-limitscalar}Expected and observed upper limits at 95\% \CL on the product of the \ttS production cross section and branching fraction at 13\TeV as a function of \mS for various relative decay widths $\GammaS/\mS$ as indicated in the legends. Data collected at 13.6\TeV are included as described in the text. The theoretical cross section for this model, as described in Section~\ref{sec:samples}, is shown as a solid blue line.}
\end{figure*}
\begin{figure*}[tb]
	\centering
	\includegraphics[width=0.42\textwidth]{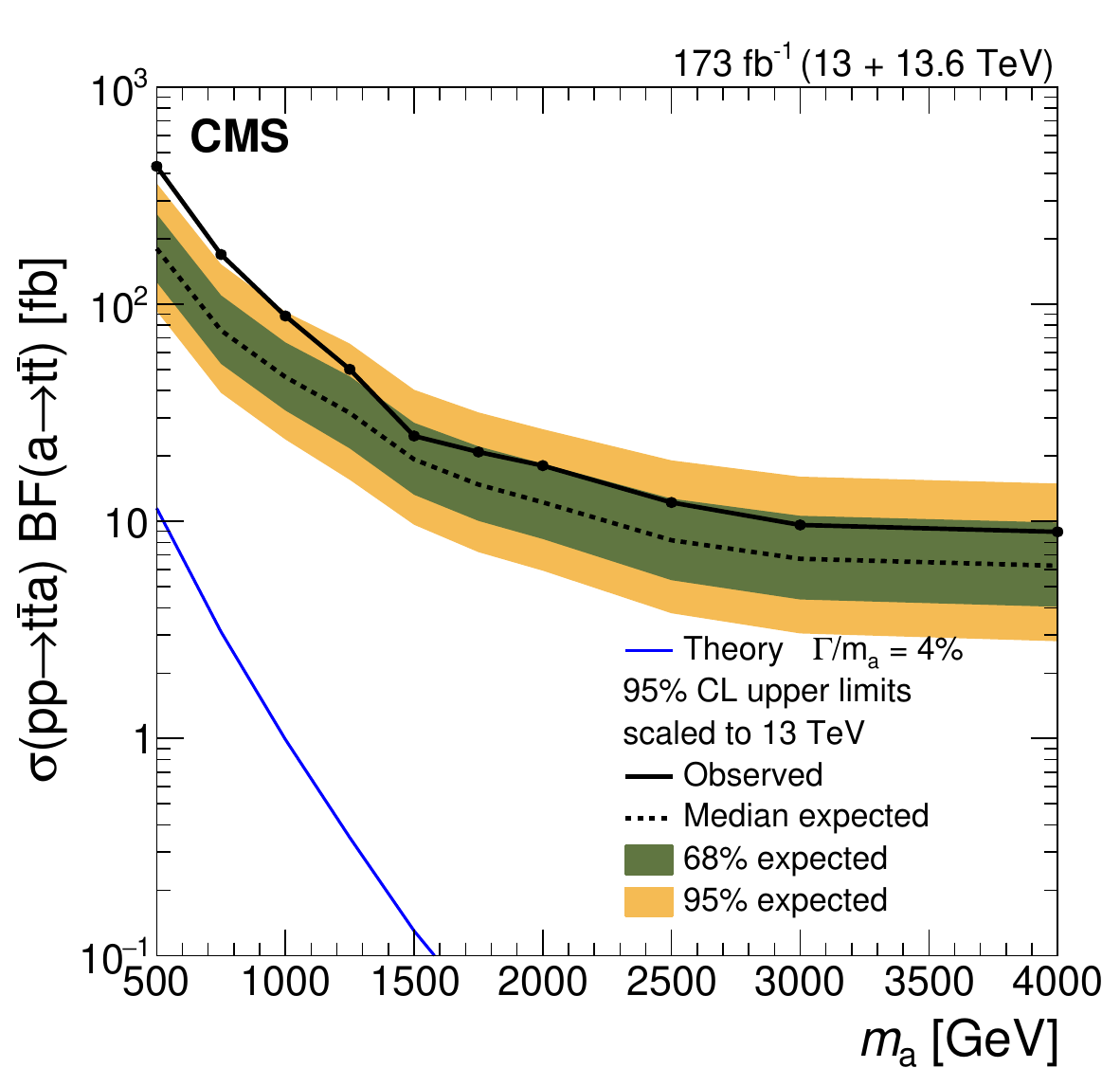}\hspace{0.03\textwidth}
	\includegraphics[width=0.42\textwidth]{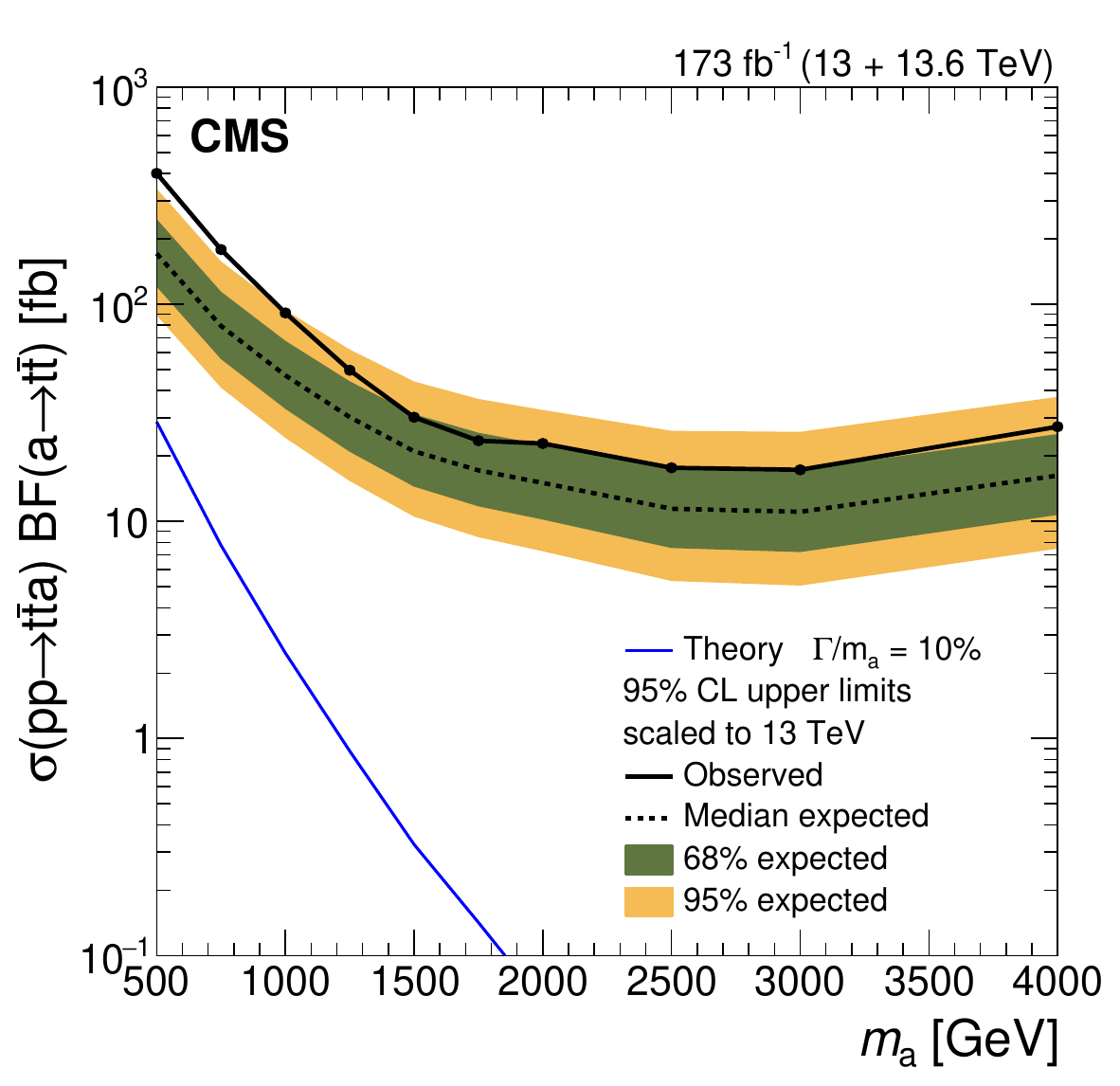}\\
	\includegraphics[width=0.42\textwidth]{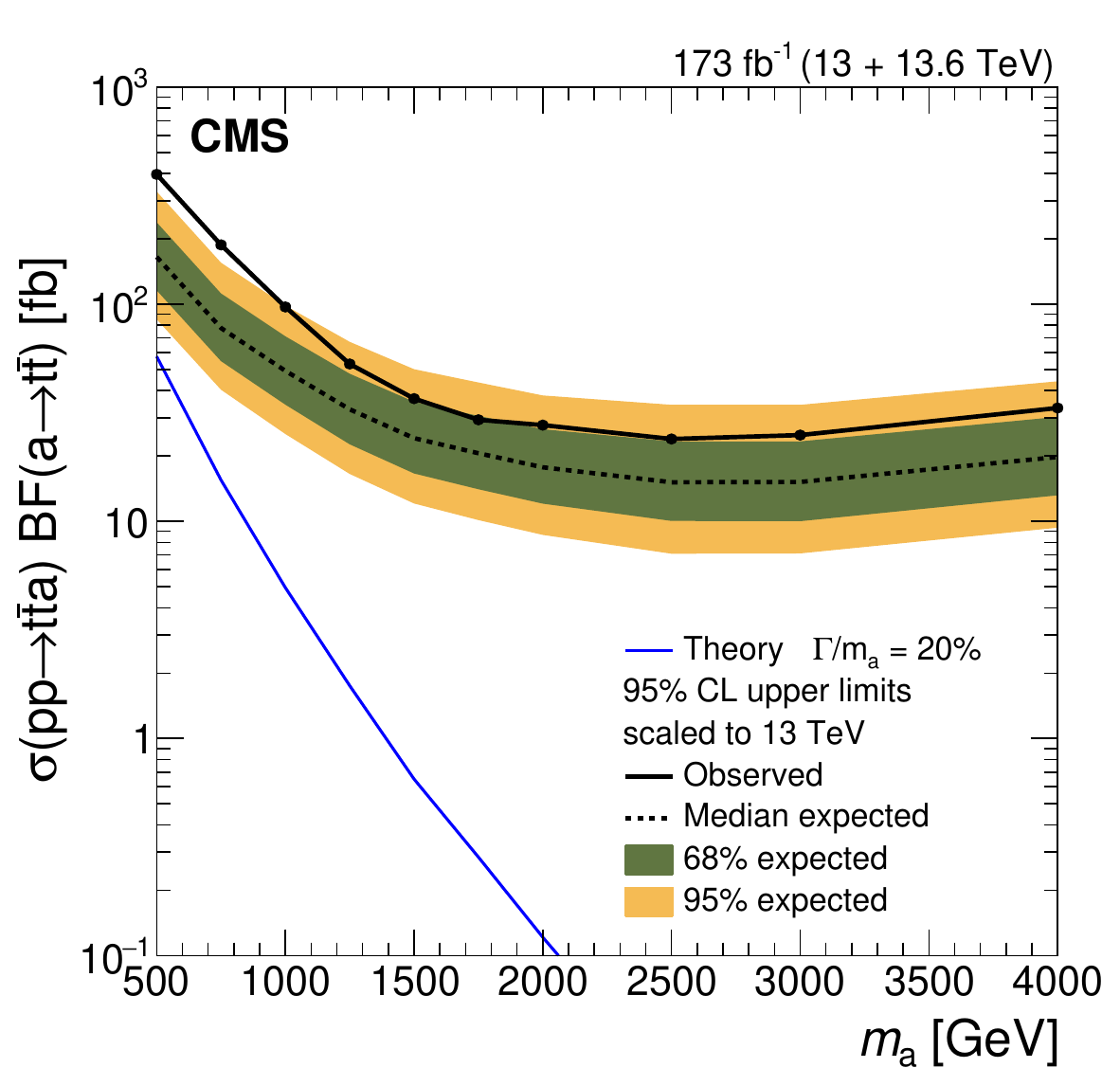}\hspace{0.03\textwidth}
	\includegraphics[width=0.42\textwidth]{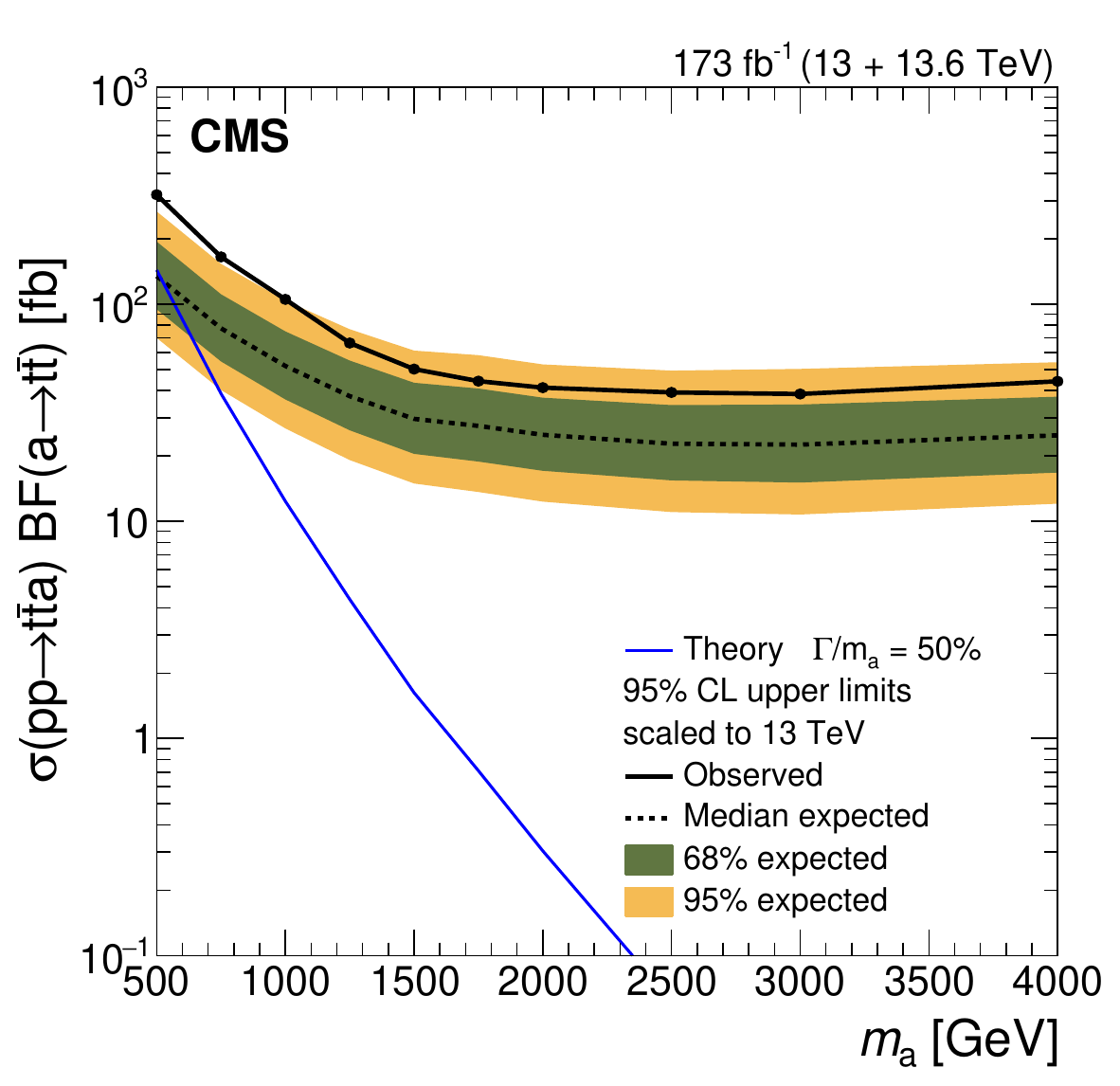}
	\caption{\label{fig:result-limitpseudoscalar}Expected and observed upper limits at 95\% \CL on the product of the \ttP production cross section and branching fraction at 13\TeV as a function of \mP for various relative decay widths $\GammaP/\mP$ as indicated in the legends. Data collected at 13.6\TeV are included as described in the text. The theoretical cross section for this model, as described in Section~\ref{sec:samples}, is shown as a solid blue line.}
\end{figure*}
Limits are also obtained for the \ttS and \ttP signal hypotheses, and are shown in Figs.~\ref{fig:result-limitscalar} and \ref{fig:result-limitpseudoscalar}, respectively. For a given mediator width, these models predict smaller cross sections than those of the \ttV model, resulting in weaker sensitivity.
Consequently, the scalar mediator hypothesis is not expected to be excluded, while the pseudoscalar model is only expected to be narrowly excluded for the 500\GeV signal with 50\% relative width. 
For these model assumptions, no signal hypothesis is excluded by the observed data. The local significance for the 500\GeV signal with 4\% relative width is 2.0 standard deviations for the \ttS process and 2.4 standard deviations for the \ttP process. These values are again consistent with statistical fluctuations of the background.

\begin{figure*}[htbp]
	\centering
	\includegraphics[width=0.7\textwidth]{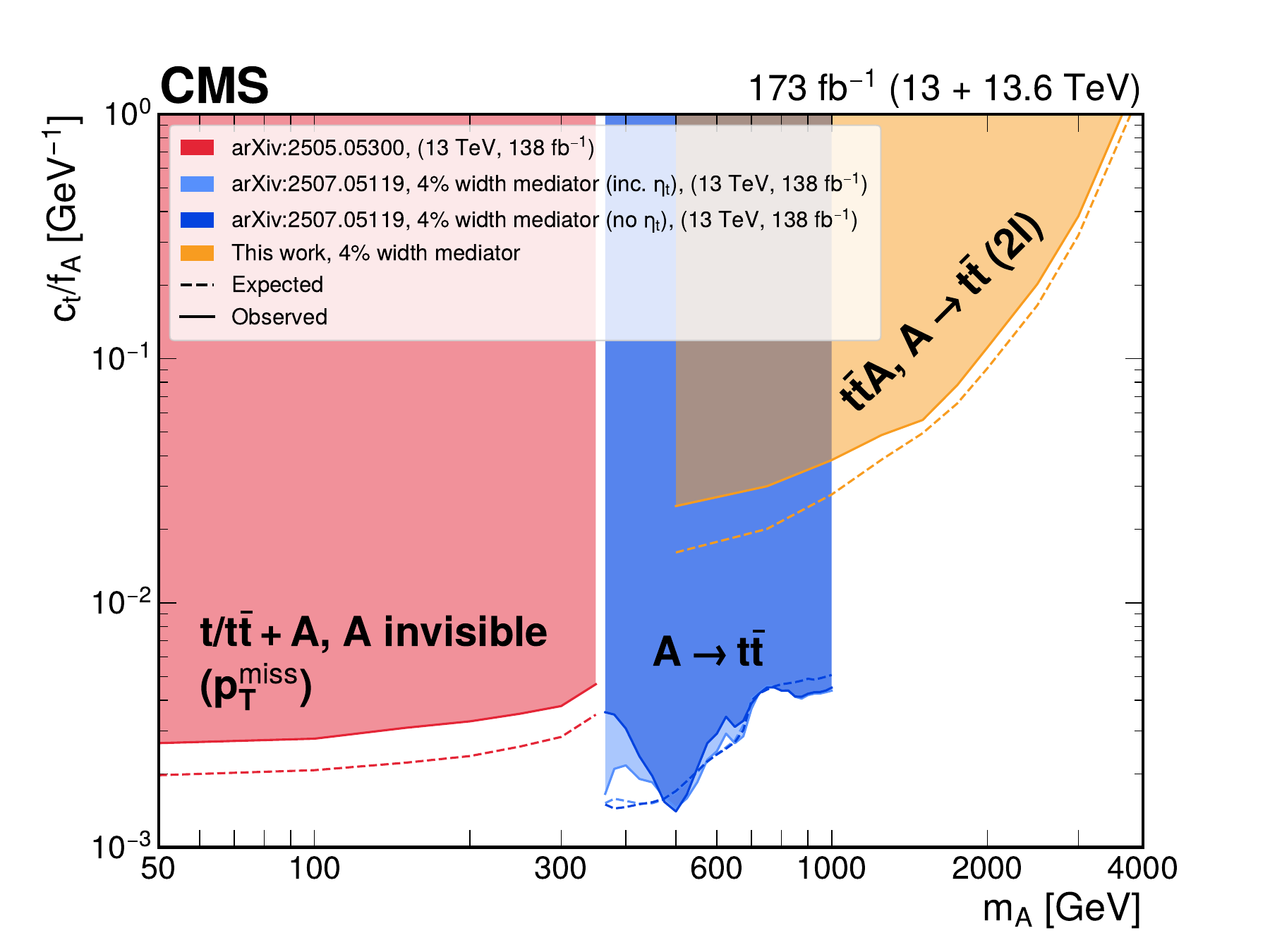}
	\caption{Expected (dashed lines) and observed (solid lines, shaded areas) upper limits at 95\% \CL on ALP-top coupling divided by the decay constant, $c_\PQt/f_{\PSA}$, from different searches: \PQt/\ttbar{}+\ptvecmiss limits from Ref.~\cite{CMS:2025ncs} (red); \ttbar resonance limits from Ref.~\cite{CMS:2025dzq} including a contribution from \etat in the background (light blue) and the limits from the same search when not considering \etat in the background (darker blue); four top quark limits from this search (orange).}
    \label{fig:result-limitALPs}
\end{figure*}

An ALP that couples to top quarks can be probed at the LHC through several complementary experimental signatures.
 Such an ALP with masses above the \ttbar threshold can be excluded by both four top quark and \ttbar resonance searches. 
An ALP with a mass below the \ttbar threshold and with a direct coupling only to the top quark, motivated by strong exclusions on couplings to other particles, is often stable on detector timescales~\cite{Esser:2023fdo}, and so can be excluded by searches for missing transverse momentum in association with top quarks. 
Each of these channels probes a different region of parameter space and relies on distinct experimental
signatures, making their comparison and combination particularly informative. For the first time in a four top quark search, the limits in Fig.~\ref{fig:result-limitpseudoscalar} are reinterpreted to provide sensitivity to ALPs from resonances in four top quark production. Four top quark final states could help distinguish an ALP at the \ttbar threshold from an additional enhancement at the \ttbar threshold predicted by non-relativistic QCD, referred to as toponium~\cite{Kiyo:2008bv, Fuks:2021xje}, since an ALP would appear in both \ttbar and four top quark final states, whereas toponium would not provide a significant enhancement in four top quark topologies. Limits on the ALP-\PQt coupling divided by the decay constant, $c_\PQt/f_{\PSA}$, are shown in Fig.~\ref{fig:result-limitALPs}.
Also shown are the exclusions from \ttbar resonance searches, taken from the CMS search for pseudoscalar resonances in $\sqrt{s}=13\TeV$ data collected in 2016--2018~\cite{CMS:2025dzq}, and a limit on ALPs below \ttbar threshold, taken from the CMS search for \ttbar and single top quark production in association with missing transverse momentum (\PQt/\ttbar{}+\ptvecmiss) in 2016--2018 data~\cite{CMS:2025ncs}.

The limits for this work are produced by converting the ratio of the exclusion limits to the theoretical cross section computed for $g_{\ttP}=1$, to limits on the coupling $g_{\ttP}$, and then using the relation in Ref.~\cite{Anuar:2024qsz} to convert this to a limit on the ALP coupling,
\begin{equation}
    \frac{c_{\PQt}}{f_{\PSA}} = \frac{g_{\ttP}}{v}.
\label{eq:result-couplingconv}
\end{equation}

The search for pseudoscalar mediators in \ttbar final states~\cite{CMS:2025dzq} uses the same model, and hence the coupling exclusions are also converted using Eq.~\eqref{eq:result-couplingconv} to produce the limits shown in Fig.~\ref{fig:result-limitALPs}. The analysis in Ref.~\cite{CMS:2025dzq} observed an excess with respect to the fixed-order background close to the \ttbar threshold which is consistent with either a BSM pseudoscalar mediator or a toponium-like enhancement. It therefore provided two sets of exclusions: one considering only the fixed order background, and another allowing for a floating contribution from a simplified toponium model, \etat; both are displayed in Fig.~\ref{fig:result-limitALPs}. Since that search only targets resolved topologies, it only sets limits up to 1000\GeV.

The exclusion for lower mediator masses can be derived using results from the \PQt/\ttbar{}+\ptvecmiss search~\cite{CMS:2025ncs}, which include a reinterpretation where the ALP is assumed to couple to both top quarks and a stable dark matter particle. Below the \ttbar{} threshold, the ALP decays almost exclusively to dark matter particles, and above the \ttbar threshold it decays to both dark matter and \ttbar. Since here we assume the ALP only couples to top quarks and is stable on detector timescales below the \ttbar threshold, we show these limits up to 350\GeV, after which a decay of the ALP to \ttbar starts to become possible. A signal with 4\% relative width is taken for the \ttbar and four top quark limits, since this is approximately the natural width for the couplings excluded by the \ttbar search~\cite{CMS:2025dzq}. The \PQt/\ttbar{}+\ptvecmiss search~\cite{CMS:2025ncs} is insensitive to the width of the mediator.

The result of the current analysis is sensitive to a higher ALP mass range because of the use of \jetJ jets. 
For an ALP with a mass of 500\GeV, couplings above $0.025\GeV^{-1}$ are excluded ($0.016\GeV^{-1}$ expected); for a mediator with a mass of 3000\GeV, the exclusion is $0.38\GeV^{-1}$ ($0.32\GeV^{-1}$ expected).

\onecolnewpage

\section{Summary}
\label{sec:summary}

A search for a heavy resonance with the CMS experiment has been presented, examining models where a new boson interacts exclusively with top quarks, resulting in events comprising four top quarks.
Two proton-proton collision data sets are analysed corresponding to 138 and 35\fbinv for centre-of-mass energies of 13 and 13.6\TeV, respectively.
The mass of the new boson is reconstructed from a pair of variable-radius jets where a boosted decision tree is used for the identification of hadronically decaying top quarks.
The additional top quarks are assumed to decay each into a \PQb quark, a lepton (electron or muon), and a neutrino, resulting in a signature with two leptons, two \PQb quark jets, and two large-radius jets.
Contributions from the dominant background processes are estimated using data from dedicated sideband regions. 

Exclusion limits on the production cross section at 95\% confidence level are determined for boson masses between 500\GeV and 4\TeV, and relative decay widths ranging between 4 and 50\% of the mediator mass.
The analysis is sensitive to scalar, pseudoscalar, and vector bosons decaying to \ttbar, and further interpretations are provided in the context of extended Higgs sectors and models with axion-like particles.
For the vector-mediator benchmark, resonances with a relative width of 50\% are excluded for masses up to 850\GeV, with an expected exclusion limit of 1000\GeV. These exclusions apply to the resonant contribution only; non-$s$-channel diagrams and interference with standard model four-top quark production are not included.
The largest deviation from the background hypothesis for the vector scenario has a local significance of 2.1 standard deviations, with similar results for the scalar and pseudoscalar mediators.
The obtained upper limits are the first results for this signature that incorporate a proton-proton collision data set at 13.6\TeV and the first dedicated search of this type in the two-lepton channel.

\begin{acknowledgments}
We congratulate our colleagues in the CERN accelerator departments for the excellent performance of the LHC and thank the technical and administrative staffs at CERN and at other CMS institutes for their contributions to the success of the CMS effort. In addition, we gratefully acknowledge the computing centres and personnel of the Worldwide LHC Computing Grid and other centres for delivering so effectively the computing infrastructure essential to our analyses. Finally, we acknowledge the enduring support for the construction and operation of the LHC, the CMS detector, and the supporting computing infrastructure provided by the following funding agencies: SC (Armenia), BMFWF and FWF (Austria); FNRS and FWO (Belgium); CNPq, CAPES, FAPERJ, FAPERGS, and FAPESP (Brazil); MES and BNSF (Bulgaria); CERN; CAS, MoST, and NSFC (China); MINCIENCIAS (Colombia); MSES and CSF (Croatia); RIF (Cyprus); SENESCYT (Ecuador); ERC PRG and PSG, TARISTU24-TK10 and MoER TK202 (Estonia); Academy of Finland, MEC, and HIP (Finland); CEA and CNRS/IN2P3 (France); SRNSF (Georgia); BMFTR, DFG, and HGF (Germany); GSRI (Greece); MATE and NKFIH (Hungary); DAE and DST (India); IPM (Iran); SFI (Ireland); INFN (Italy); MSIT and NRF (Republic of Korea); MES (Latvia); LMTLT (Lithuania); MOE and UM (Malaysia); BUAP, CINVESTAV, CONACYT, LNS, SEP, and UASLP-FAI (Mexico); MOS (Montenegro); MBIE (New Zealand); PAEC (Pakistan); MSHE, NSC, and NAWA (Poland); FCT (Portugal); MESTD (Serbia); MICIU/AEI and PCTI (Spain); MOSTR (Sri Lanka); Swiss Funding Agencies (Switzerland); MST (Taipei); MHESI (Thailand); TUBITAK and TENMAK (T\"{u}rkiye); NASU (Ukraine); STFC (United Kingdom); DOE and NSF (USA).

\begin{sloppypar}
\setlength\emergencystretch{\hsize}
\hyphenation{Rachada-pisek} Individuals have received support from the Marie-Curie programme and the European Research Council and Horizon 2020 Grant, contract Nos.\ 675440, 724704, 752730, 758316, 765710, 824093, 101115353, 101002207, 101001205, and COST Action CA16108 (European Union); the Leventis Foundation; the Alfred P.\ Sloan Foundation; the Alexander von Humboldt Foundation; the Science Committee, project no. 22rl-037 (Armenia); the Fonds pour la Formation \`a la Recherche dans l'Industrie et dans l'Agriculture (FRIA) and Fonds voor Wetenschappelijk Onderzoek contract No. 1228724N (Belgium); the Beijing Municipal Science \& Technology Commission, No. Z191100007219010, the Fundamental Research Funds for the Central Universities, the Ministry of Science and Technology of China under Grant No. 2023YFA1605804, the Natural Science Foundation of China under Grant No. 12535004, and USTC Research Funds of the Double First-Class Initiative No.\ YD2030002017 (China); the Ministry of Education, Youth and Sports (MEYS) of the Czech Republic; the Shota Rustaveli National Science Foundation (Georgia); the Deutsche Forschungsgemeinschaft (DFG), among others, under Germany's Excellence Strategy -- EXC 2121 ``Quantum Universe" -- 390833306, and under project number 400140256 - GRK2497; the Hellenic Foundation for Research and Innovation (HFRI), Project Number 2288 (Greece); the Hungarian Academy of Sciences, the New National Excellence Program - \'UNKP, the NKFIH research grants K 131991, K 138136, K 143460, K 143477, K 147557, K 146913, K 146914, K 147048, TKP2021-NKTA-64, and 2025-1.1.5-NEMZ\_KI-2025-00004, and MATE KKP and KKPCs Research Excellence and Flagship Research Groups grants (Hungary); the Council of Science and Industrial Research, India; ICSC -- National Research Centre for High Performance Computing, Big Data and Quantum Computing, FAIR -- Future Artificial Intelligence Research, and CUP I53D23001070006 (Mission 4 Component 1), funded by the NextGenerationEU program, the Italian Ministry of University and Research (MUR) under Bando PRIN 2022 -- CUP I53C24002390006, PRIN PRIMULA 2022RBYK7T (Italy); the Latvian Council of Science; the Ministry of Science and Higher Education, project no. 2022/WK/14, and the National Science Centre, contracts Opus 2021/41/B/ST2/01369, 2021/43/B/ST2/01552, 2023/49/B/ST2/03273, and the NAWA contract BPN/PPO/2021/1/00011 (Poland); the Funda\c{c}\~ao para a Ci\^encia e a Tecnologia (Portugal); the National Priorities Research Program by Qatar National Research Fund;  MICIU/AEI/10.13039/501100011033, ERDF/EU, ``European Union NextGenerationEU/PRTR", projects PID2022-142604OB-C21, PID2022-139519OB-C21, PID2023-147706NB-I00, PID2023-148896NB-I00, PID2023-146983NB-I00, PID2023-147115NB-I00, PID2023-148418NB-C41, PID2023-148418NB-C42, PID2023-148418NB-C43, PID2023-148418NB-C44, PID2024-158190NB-C22, RYC2021-033305-I, RYC2024-048719-I, CNS2023-144781, CNS2024-154769 and Plan de Ciencia, Tecnolog{\'i}a e Innovaci{\'o}n de Asturias, Spain; the Chulalongkorn Academic into Its 2nd Century Project Advancement Project, the National Science, Research and Innovation Fund program IND\_FF\_68\_369\_2300\_097, and the Program Management Unit for Human Resources \& Institutional Development, Research and Innovation, grant B39G680009 (Thailand); the Eric \& Wendy Schmidt Fund for Strategic Innovation through the CERN Next Generation Triggers project under grant agreement number SIF-2023-004; the Kavli Foundation; the Nvidia Corporation; the SuperMicro Corporation; the Welch Foundation, contract C-1845; and the Weston Havens Foundation (USA).
\end{sloppypar}
\end{acknowledgments}\section*{Data availability} Release and preservation of data used by the CMS Collaboration as the basis for publications is guided by the  \href{https://doi.org/10.7483/OPENDATA.CMS.1BNU.8V1W}{CMS data preservation, re-use and open access policy}.

\bibliography{auto_generated}

\cleardoublepage \appendix\section{The CMS Collaboration \label{app:collab}}\begin{sloppypar}\hyphenpenalty=5000\widowpenalty=500\clubpenalty=5000\cmsinstitute{Yerevan Physics Institute, Yerevan, Armenia}
{\tolerance=6000
A.~Gevorgyan\cmsorcid{0000-0003-2751-9489}, A.~Hayrapetyan, V.~Makarenko\cmsorcid{0000-0002-8406-8605}, A.~Tumasyan\cmsAuthorMark{1}\cmsorcid{0009-0000-0684-6742}
\par}
\cmsinstitute{Marietta Blau Institute for Particle Physics, Vienna, Austria}
{\tolerance=6000
W.~Adam\cmsorcid{0000-0001-9099-4341}, L.~Benato\cmsorcid{0000-0001-5135-7489}, T.~Bergauer\cmsorcid{0000-0002-5786-0293}, M.~Dragicevic\cmsorcid{0000-0003-1967-6783}, S.~Gundacker\cmsorcid{0000-0003-2087-3266}, A.K.~Guven\cmsorcid{0009-0004-5670-5138}, P.S.~Hussain\cmsorcid{0000-0002-4825-5278}, M.~Jeitler\cmsAuthorMark{2}\cmsorcid{0000-0002-5141-9560}, N.~Krammer\cmsorcid{0000-0002-0548-0985}, A.~Li\cmsorcid{0000-0002-4547-116X}, D.~Liko\cmsorcid{0000-0002-3380-473X}, M.~Matthewman, J.~Schieck\cmsAuthorMark{2}\cmsorcid{0000-0002-1058-8093}, R.~Sch\"{o}fbeck\cmsAuthorMark{2}\cmsorcid{0000-0002-2332-8784}, M.~Shooshtari\cmsorcid{0009-0004-8882-4887}, M.~Sonawane\cmsorcid{0000-0003-0510-7010}, N.~Van~Den~Bossche\cmsorcid{0000-0003-2973-4991}, W.~Waltenberger\cmsorcid{0000-0002-6215-7228}, C.E.~Wulz\cmsAuthorMark{2}\cmsorcid{0000-0001-9226-5812}
% ...
\par}
$^{1}$Also at Yerevan State University, Yerevan, Armenia\\
$^{2}$Also at Technische Universit\"{a}t Wien, Vienna, Austria\\
% ...
\end{sloppypar}
%%% END EDITABLE REGION %%%
% skeleton_end
\end{document}